# Provably Fair Cooperative Scheduling

## Reiner Hähnle[a] and Ludovic Henrio[b]

a   Department of Computer Science, TU Darmstadt, Germany
b   Univ Lyon, EnsL, UCBL, CNRS, Inria, LIP, Lyon, France

**Abstract**   The *context* of this work is cooperative scheduling, a concurrency paradigm, where task execution is not arbitrarily preempted. Instead, language constructs exist that let a task voluntarily yield the right to execute to another task.

The *inquiry* is the design of provably fair schedulers and suitable notions of fairness for cooperative scheduling languages. To the best of our knowledge, this problem has not been addressed so far.

Our *approach* is to study fairness independently from syntactic constructs or environments, purely from the point of view of the semantics of programming languages, i.e., we consider fairness criteria using the formal definition of a program execution. We develop our concepts for classic structural operational semantics (SOS) as well as for the recent *locally abstract, globally concrete* (LAGC) semantics. The latter is a highly modular approach to semantics ensuring the separation of concerns between local statement evaluation and scheduling decisions.

The new *knowledge* contributed by our work is threefold: first, we show that a new fairness notion, called *quiescent* fairness, is needed to characterize fairness adequately in the context of cooperative scheduling; second, we define a provably fair scheduler for cooperative scheduling languages; third, a qualitative comparison between the SOS and LAGC versions yields that the latter, while taking higher initial effort, is more amenable to proving fairness and scales better under language extensions than SOS.

The *grounding* of our work is a detailed formal proof of quiescent fairness for the scheduler defined in LAGC semantics.

The *importance* of our work is that it provides a formal foundation for the implementation of fair schedulers for cooperative scheduling languages, an increasingly popular paradigm (for example: akka/Scala, JavaScript, async Rust). Being based solely on semantics, our ideas are widely applicable. Further, our work makes clear that the standard notion of fairness in concurrent languages needs to be adapted for cooperative scheduling and, more generally, for any language that combines atomic execution sequences with some form of preemption.

# The Art, Science, and Engineering of Programming

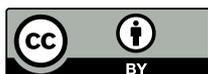

## 1 Introduction

In this article we study, from a semantic point of view, the design of fair schedulers, in particular, for languages with cooperative scheduling. Cooperative scheduling denotes a concurrency paradigm, where task execution is not arbitrarily preempted. Instead, there are language constructs letting a task voluntarily yield control (hence the adjective *cooperative*). Cooperative scheduling is implemented in active object languages [6] and recently became more widespread (for example, in akka/Scala, JavaScript, async Rust).

Specifically, our motivation to study fairness [17] comes from the challenge to implement the active object language ABS [20]. Which scheduling policy should the ABS runtime system [28] choose? Existing notions of (weak) fairness were clearly inapplicable. The pragmatic solution is to realize a non-deterministic scheduler that randomly selects an enabled process with equal probability [2] , conjectured to be asymptotically fair, however, the absence of an adequate fairness notion and an at least weakly fair deterministic scheduler is clearly unsatisfactory. We remedy this situation in the present article in a sufficiently abstract manner for our results to be applicable to any language with cooperative scheduling.

Generally speaking, fairness properties state that "considering an event that could occur infinitely often, this event will occur eventually". In particular, fairness provides no guarantee for terminating programs or stuck processes. In distributed systems, certain fairness notions quantify over the number of scheduled processes before an event occurs: This is not what we study here.

First, we study fairness independently from dedicated syntactic constructs or environments, purely from the point of view of the semantics of programming languages, i.e., we consider fairness criteria using the formal definition of a program execution. Specifically, we compare a classic structural operational semantics (SOS) with the more recent *locally abstract, globally concrete* (LAGC) semantics [12, 13]. LAGC is a highly modular approach to semantics ensuring the separation of concerns between local statement evaluation and scheduling decisions. In either case, different notions of fairness are based on the principle that *enabled actions* are the *reachable statements*.

Second, we explain that the traditional weak fairness criterion is inadequate in the context of cooperative scheduling, because it relies on actions that can be continuously triggered. This is not applicable in cooperative scheduling, where a process runs uninterrupted until the next yield point. We detail and formalize this issue in Section 3.3 and suggest an alternative, more adequate, notion of weak fairness based on so-called *quiescent* states. *Quiescent fairness* amounts to checking the fairness condition only at quiescent states, i.e. at (cooperative) scheduling points. Incidentally, standard weak fairness is inadequate not only for languages with cooperative scheduling, but for any language that combines *atomically executed sequences* with some form of preemption, such as ProMeLa [19].

Third, we define a quiescent fair scheduler for languages with cooperative scheduling in SOS as well as LAGC semantics. For the minimalist language in this paper, the LAGC definitions are slightly more complex, but also more modular than SOS, because they strictly decouple the local semantics of individual statements from scheduling

$$s \in Stmt ::= \mathsf{skip} \mid v := e \mid \mathsf{if}\ e\ \{\,s\,\} \mid s;s \mid \mathsf{while}\ e\ \{\,s\,\}$$

■ **Figure 1** The syntax for statements in While.

$$P \in Prog ::= \overline{M}\ \{s\} \qquad\qquad s \in Stmt ::= s_0 \mid \mathsf{spawn}(m,e)$$
$$M \in Proc ::= m(x)\{s\}$$

■ **Figure 2** Program syntax with procedure calls, $s_0$ are the statements introduced in Figure 1.

and composing them into a global trace. This modularity permits to *prove* quiescent fairness of the LAGC scheduler in a reasonably compact manner; such a proof would be much more cumbersome with the SOS version. To the best of our knowledge, we provide the first scheduler for cooperative scheduling proven to be fair.

To make the paper more accessible, we develop our notions along a sequence of languages of increasingly complexity. These are defined in Section 2. Then Section 3 discusses fair scheduling for cooperative scheduling languages in the context of SOS. As LAGC semantics is less known, we give its essentials in Section 4 to make the paper self-contained. Next we develop and *prove* fairness of a scheduler in three successive stages in Sects. 5, 6, 7, where the latter addresses cooperative scheduling. Section 8 compares the SOS and LAGC semantics with respect to the design of fair schedulers. Section 9 discusses work related both to fairness and to different ways to write programming language semantics; we conclude in Section 10. Proofs and auxiliary notions are located in the appendix to improve readability.

## 2 Languages

In this paper we consider four languages of increasing complexity:

1. A standard sequential language called While;
2. An extension of While with concurrent procedures called Spawn;
3. An extension of Spawn with guarded statements;
4. The cooperative scheduling language CoopWhile, containing spawn, suspend,[1] and join.

We collect the definitions of these languages in the present section for future reference. Figure 1 contains the grammar of While. There is a fixed set of global variables $v$. We leave out the expression syntax *Exp* and suppose that expressions are side effect-free and well-typed so that expression evaluation does not get stuck. Standard Boolean and arithmetic expressions appear in the examples.

Figure 2 defines an extension of While with a spawn statement: A program $P$ is a set of procedure declarations $\overline{M}$ and a main block $\{s\}$. A procedure declaration associates a name $m$ with the statement in its body. To reduce technicalities, we consider procedures with one local parameter and no return value. We call this language Spawn.

---

[1] In some languages this statement is called "yield".

$$s \in Stmt ::= s_0 \mid \mathsf{spawn}(m, e, x) \mid \mathsf{suspend} \mid \mathsf{await}\ x \mid \mathsf{return}$$

**Figure 3** Syntax of CoopWhile: $s_0$ are statements of While in Figure 1, procedure calls and programs are defined as for Spawn in Figure 2. The return statement only appears at *runtime*.

Guarded statements are a common synchronization mechanism in concurrent languages [10, 19, 20]. A statement is preceded by a Boolean guard expression that blocks execution of the current process, until the guard is evaluated to true. This can be used, for example, to ensure that the result of a computation is ready before it is used, a message has arrived, etc. Our syntax for guarded commands is inspired by PROMELA [19]; in the Guard language the syntax for statements from Figure 2 becomes:

$$s \in Stmt ::= (:: g; s) \mid s_1 \ ,$$

where $g$ is a Boolean expression and $s_1$ are the statements defined in Figure 2.

Finally, we define a minimal language with spawn, suspend, and join that nevertheless contains all complications of cooperative scheduling called CoopWhile. The syntax of the language CoopWhile is shown in Figure 3. Each spawn statement creates a task identifier stored in $x$ that can later be waited upon with await $x$ (we assume task identifiers are initialized before they are used). It is also possible that a task voluntarily yields the processor to other tasks with the unconditional suspend command.

## 3 Fair Scheduling Based on Structural Operational Semantics

We explain how to model fairness in a straightforward, classical SOS semantics setting. As SOS is very common, we jump directly to the CoopWhile language that raises interesting questions in presence of cooperative scheduling. The semantics presented in Section 3.1 is pretty standard and could be skipped by the knowledgeable reader, however, we use it in Section 3.3 to explain the principles of our fairness criteria in an SOS setting.

### 3.1 SOS for CoopWhile

We begin with an SOS semantics for CoopWhile, as an extension of the standard SOS semantics for While [22]. We assume for While a binary relation $(s, \sigma) \xrightarrow{\text{While}} (s', \sigma')$ between pairs of statements $s$ and computation states $\sigma$ (mapping variables to values), such that $s$ is reduced to $s'$ in one step while $\sigma$ is updated to $\sigma'$. As usual, we rely on a function $\text{val}_\sigma(e)$ for evaluating expressions, parameterized by $\sigma$.

The SOS rules for statements in CoopWhile that are not part of While, are in Figure 4. Configurations are either of the form $f \cdot s$, $T$, $\sigma$, when exactly one task with identifier $f$ and current statement $s$ is active, or Idle, $T$, $\sigma$ when no task is currently active. $T$ is a mapping from task identifiers to suspended tasks, $\sigma$ represents shared memory. The rules are inspired by semantics for active object and cooperative languages [1, 7, 20].

$$\frac{(s, \sigma) \xrightarrow{\text{while}} (s', \sigma')}{f \cdot s, T, \sigma \to f \cdot s', T, \sigma'} \text{ Local} \qquad \qquad \begin{array}{l} \text{Return} \\ f \cdot \text{return}, T, \sigma \to \text{Idle}, T, \sigma \end{array}$$

SpawnStart
$$\frac{f' \notin \text{dom}(T) \cup \{f\} \qquad m(z)\{s'\} \in \overline{M} \qquad y \notin \text{dom}(\sigma)}{f \cdot \text{spawn}(m, e, x); s, T, \sigma \to f \cdot s, T[f' \mapsto s'[z \leftarrow y]; \text{return}], \sigma[y \mapsto \text{val}_\sigma(e), x \mapsto f']}$$

YieldSuspend
$f \cdot \text{suspend}; s, T, \sigma \to \text{Idle}, T[f \mapsto s], \sigma$

YieldAwait
$f \cdot \text{await } x; s, T, \sigma \to \text{Idle}, T[f \mapsto \text{await } x; s], \sigma$

ScheduleSimple
$$\frac{s \text{ is not of the form await } x; s'}{\text{Idle}, T \uplus [f \mapsto s], \sigma \to f \cdot s, T, \sigma}$$

ScheduleAwait
$$\frac{s = \text{await } x; s' \qquad \text{val}_\sigma(x) \notin \text{dom}(T)}{\text{Idle}, T \uplus [f \mapsto s], \sigma \to f \cdot s', T, \sigma}$$

■ **Figure 4** SOS semantics for `CoopWhile`

Rule Local triggers the SOS rules for `While` for an active task $f$. Rule SpawnStart spawns a task, associates a fresh task identifier $f'$ to it, and assigns it to $x$. It also looks up the declaration of $m$, renames its parameter $z$ to a fresh $y$; the new task associates $f'$ with the renamed procedure body. Then rule SpawnStart continues with the trailing statement $s$ after the procedure call. The end of the spawned task is marked with statement `return` which is only used for this purpose. Its SOS rule Return releases the processor. Both `suspend` and `await` release the processor, the difference is that the task identifier $x$ of `await` needs to be queried and is thus kept. Scheduling takes place whenever there is no active ask and is performed by the two scheduling rules. Again, the difference is whether a wait condition must be checked, indeed an await statement can only be scheduled if the awaited task is finished.

To evaluate a `CoopWhile` program with main block $s_P$, one must start from an *initial configuration* of the form $f \cdot s_P, \emptyset, \emptyset$ for an arbitrary $f$.

### 3.2 SOS of `CoopWhile` with Scheduling

In Figure 5 we add scheduling to the SOS of `CoopWhile`, i.e. we resolve the non-determinism in the scheduling rules (all other rules are deterministic). To this end we let $q$ range over queues of pairs of task identifiers $f$ and statements $s$. Queues are equipped with the following operations: create() creates an empty queue, push($q, k$) returns $q$ extended with $k$ as its final element; pop($q$) returns a pair ($q', k$), where $k$ is the first element in $q$ and $q'$ the rest of the queue, rotate($q$) shifts the beginning of the queue by one element, turning the new queue (App. A formally defines queues).

Rules Local and Return are unchanged except replacing $f$ with $q$ and we omit them. Rule SpawnStart pushes the new task to the end of $q$, instead of extending mapping $f$, and is unchanged otherwise. The same holds for the Yield rules, which we omit as well. The only non-trivial change is required for scheduling: One always

SPAWNSTART

$$\frac{f' \notin \mathrm{dom}(T) \cup \{f\} \qquad m(z)\{s\} \in \overline{M} \qquad y \notin \mathrm{dom}(\sigma)}{f \cdot \mathrm{spawn}(m, e, x); s, q, \sigma \to f \cdot s, \mathrm{push}\big(q, (f', s[z \leftarrow y]; \mathrm{return})\big), \sigma[y \mapsto \mathrm{val}_\sigma(e), x \mapsto f']}$$

$$\frac{\mathrm{pop}(q) = \big(q', (f, s)\big) \qquad s \text{ is not of the form } \mathsf{await}\ x; s'}{\mathrm{Idle}, q, \sigma \to f \cdot s, q', \sigma} \ \text{SCHEDULESIMPLE}$$

$$\frac{\mathrm{pop}(q) = \big(q', (f, s)\big) \qquad s = \mathsf{await}\ x; s' \qquad \mathrm{val}_\sigma(x) \notin \mathrm{Tsk}(q)}{\mathrm{Idle}, q, \sigma \to f \cdot s', q', \sigma} \ \text{SCHEDULEAWAITDONE}$$

$$\frac{\mathrm{pop}(q) = \big(q', (f, s)\big) \qquad s = \mathsf{await}\ x; s' \qquad \mathrm{val}_\sigma(x) \in \mathrm{Tsk}(q)}{\mathrm{Idle}, q, \sigma \to \mathrm{Idle}, \mathrm{rotate}(q), \sigma} \ \text{SCHEDULEAWAITWAIT}$$

■ **Figure 5** SOS scheduler semantics for `CoopWhile`, where $\mathrm{Tsk}(q) = \{f \mid \exists s. (f, s) \in q\}$

(re-)activates the first task in $q$ by retrieving it with pop, but this does not succeed for an `await` whose task identifier $x$ has not finished. In that case, the queue is rotated in rule SCHEDULEAWAITWAIT. For a `CoopWhile` program with main block $s_P$, the *initial configuration* is $f \cdot s_P$, create(), $\emptyset$ for some $f$.

### 3.3 Fairness

We design now a notion of fairness that is inherently tied to program semantics, not the programs themselves or a specific interpreter. First, we label SOS transitions with task identifiers to record which task progresses, but this is very operational, so in a second version we state an alternative fairness criterion based on *reachable statements*. It is important to note that, by definition, both weak and strong fairness conditions apply only if a considered event can be scheduled regularly in the future. This implies that when a process never yields and does not terminate, then the underlying program never *reaches* a scheduling point, hence, the single possible execution that always runs the same thread is "fair" by definition.

#### 3.3.1 Using Task Identifiers to Define Fairness

We assume that every SOS rule application is labeled with the task identifier that progresses (or with the identifier of the task being suspended). Transitions are then of the form $C \xrightarrow{f} C'$ with $C$ and $C'$ the source and target SOS configuration, respectively, and $f$ the task identifier of the task that progresses. We define enabled($C$) as the set of task identifiers that can progress: enabled($C$) = $\{f \mid \exists C'. C \xrightarrow{f} C'\}$. We can then express standard fairness criteria as follows:

**Definition 1** (Fairness). *Given an infinite sequence of SOS rule applications $C_0 \xrightarrow{f_1} C_1 \xrightarrow{f_2} \cdots \xrightarrow{f_n} C_n \cdots$, we say that the resulting execution is* weakly fair *if:*

$$\forall m, f. \left( (\forall n \geq m. (f \in \text{enabled}(C_n))) \Rightarrow \left( \exists n' \geq m. \left( C_{n'} \xrightarrow{f} C_{n'+1} \right) \right) \right) .$$

*We say that the resulting execution is* strongly fair *if:*

$$\forall m, f. \left( (\forall n \geq m. \exists l \geq n. (f \in \text{enabled}(C_l))) \Rightarrow \left( \exists n' \geq m. \left( C_{n'} \xrightarrow{f} C_{n'+1} \right) \right) \right) .$$

Weak fairness requires a statement to be always "schedulable" from a certain point onward to guarantee that it will be scheduled; in contrast, strong fairness merely requires the statement to be "schedulable" infinitely often.

### 3.3.2 A Fairness Criterion for Cooperative Scheduling

The standard notion of weak fairness as defined above is too weak to be useful for cooperative scheduling. As soon as one task is active, the other tasks cannot be scheduled immediately, which makes the precondition for weak fairness inapplicable. In our setting, the only point where several different tasks belong to the set of enabled actions is when no task is currently active. When a task is scheduled the other task identifiers are *not* enabled any more. Thus, weak fairness provides no guarantee in a cooperative scheduling context, as shown by the Example 1 below. Regarding strong fairness, it is generally impossible to design a deterministic strongly fair scheduler valid for any program, because, in `CoopWhile` and in many languages, strong fairness can only be achieved probabilistically. See Example 8 for a discussion on this point.

**Example 1** (Inadequacy of weak fairness for cooperative scheduling). *Consider the `CoopWhile` program $P$ with main block "$\{$`spawn`$(m, 0, z)$; `suspend`; $j := 2\}$" and a procedure $m$ declared as "$m(x)\{$`while true` $\{$`suspend`; $i := 1\}\}$". We ask the following question: Does fairness ensure that statement "$j := 2$" will be executed at some point? Strong fairness would be sufficient, but it is too difficult to implement. In contrast, weak fairness is insufficient, because the execution that suspends the main thread and then always schedules the spawned task is weakly fair. Every time the `while` statement is reached, the spawned task is active and the main thread is unable to progress. Weak fairness merely requires actions to be* always schedulable *to progress, so it is insufficient to ensure that "$j := 2$" is executed.*

*Instead, if we only check schedulability at* suspension points*, we note that "$j := 2$" can be executed each time the spawned thread is suspended. Our new notion of fairness defined below, called* quiescent *fairness, ensures that "$j := 2$" will be executed.*

We define a notion of weak fairness adequate for cooperative scheduling which, instead of demanding that a statement can be scheduled at *any stage $n \geq m$*, merely stipulates this needs to be the case whenever there is *no currently running task*. The precondition then becomes "from a certain point on in the execution, *whenever the configuration is idle*, the considered task can be scheduled". We call idle configurations *quiescent states* and derive a corresponding fairness notion called *quiescent fairness*. A related concept in the context of I/O automata has been suggested in [26, 27].

**Definition 2** (Quiescent Fairness)**.** *Let* quiescent$(C)$ *be true if* $C$ *is of the form* Idle, $T$, $\sigma$. *An execution is* quiescent fair *if a statement that can be scheduled at each quiescent state after a certain point is eventually scheduled:*

$$\forall m, f. \left( \forall n > m. \, (\text{quiescent}(C_n) \Rightarrow f \in \text{enabled}(C_n)) \Rightarrow \exists n' > m. \left( C_{n'} \xrightarrow{f} C_{n'+1} \right) \right) \, .$$

Quiescent fairness is sufficient to ensure that a task that can progress will eventually be scheduled, even if another task is permanently blocked, e.g. on an await statement. Quiescent fairness is only interesting for programs that have no infinite local computation. We restrict ourselves to these programs in the following, formally:

**Definition 3** (No Infinite Local Computation)**.** *A program has* no infinite local computation *if at any stage m of an execution starting from the initial program state there is a following stage $n > m$ that is a quiescent state.*

**Example 2.** *The program in Example 1 has no infinite local computations, thus a quiescent fairness will ensure that "$j := 2$" is executed.*

### 3.3.3 Fairness Based on Reachable Statements

Task identifiers are an operational notion and reasoning over them is restrictive. For example, a task might progress infinitely often, but never when it can enter a relevant branch, and this is not observed in Def. 1. An alternative is to use the set of *reachable* statements, called newStmt$(C, C')$, defined as the set of statements introduced by a single reduction step from $C$ to $C'$. It represents the computational state of the threads that made progress during that reduction.

**Definition 4** (Set of Possible Continuations)**.** *Given two configurations $C$, $C'$ such that $C \rightarrow C'$, we define* newStmt$(C, C')$ *as follows:*

$$\begin{aligned}
\text{newStmt}((f \cdot s, T, \sigma), (f \cdot s', T', \sigma')) &= \{s'\} \cup \{s'' \mid \exists f'. \, T' = T \cup [f' \mapsto s'']\} \\
\text{newStmt}((f \cdot s, T, \sigma), (Idle, T', \sigma')) &= \emptyset
\end{aligned}$$

enabled$(C)$ *is the multiset of new statements reachable from $C$:*

$$\text{enabled}(C) = \{\text{newStmt}(C, C') \mid \exists C'. \, C \rightarrow C'\}$$

We observe that enabled$(C)$ is an infinite set, whenever a fresh name or identifier is created, but this has no adverse consequences on the definition of fairness (it is also possible to render this set finite). It is a multiset, because identical statements may result from different rule applications. This way to define enabled actions is more abstract than in Section 3.3.1 and it is *less* precise in certain cases, because different tasks can result in identical statements. It also raises the question of statements that are reduced to themselves: if $(s, \sigma) \rightarrow (s, \sigma)$ is a valid reduction of the local semantics, is $s$ an enabled statement or not? However, identical statements that really originate from different tasks can always be syntactically distinguished, for example, by their position in the code, or even using a task identifier in addition to the statement for even more precision.

Now we adapt weak, strong, and quiescent fairness to the new version of enabled($c$):

**Definition 5** (Fairness based on reachable statements)**.** *Given an infinite sequence of SOS rule applications $C_0 \to C_1 \to \cdots \to C_n \cdots$, the resulting execution is* weakly fair *if:*

$$\forall m, s. \left( (\forall n \geq m. (s \in \text{enabled}(C_n))) \Rightarrow \left( \exists n' \geq m. (s \in \text{newStmt}(C_{n'}, C_{n'+1})) \right) \right) .$$

*It is* strongly fair *if:*

$$\forall m, s. \left( (\forall n \geq m. \exists l \geq n. (s \in \text{enabled}(C_l))) \Rightarrow \left( \exists n' \geq m. (s \in \text{newStmt}(C_{n'}, C_{n'+1})) \right) \right) .$$

*It is* quiescent fair *if:*

$$\forall m, s. \left( \forall n > m. (\text{quiescent}(C_n) \Rightarrow s \in \text{enabled}(C_n)) \right.$$
$$\left. \Rightarrow \exists n' > m. s \in \text{newStmt}(C_{n'}, C_{n'+1}) \right)$$

The main drawback of the reachable statement approach is that the definition of newStmt is somewhat *ad hoc* and dependent on the structure of the configuration. Also the interplay between the definition of configurations and their evolution according to the operational semantics changes depending on the features of the language. LAGC semantics will provide us with a more general and modular solution.

## 4 LAGC Semantics

We now give an alternative semantics called LAGC to the languages in Section 2. It requires somewhat more technical apparatus than SOS: first, the local semantics of sequential code is *parameterized* over a context; second, the evaluation of some statements creates *events*; third, we use *well-formedness* predicates ensuring that instantiation and sequencing of events results in the intended semantics. While this setup is somewhat more complex than SOS, it has two advantages: it completely separates rules that *progress*, i.e. evaluate a statement, from the so-called *composition* rules responsible for making a scheduling decision. Since the composition rules are formulated without referring to specific statements, one can obtain different concurrent semantics simply by varying the scheduling rules and the well-formedness constraints, while everything else stays the same. In particular, to introduce scheduling, one needs to change only a small number of rules.

The main principle of LAGC semantics is to strictly separate two phases. The first phase evaluates statements to sets of local traces with parameters. The second phase composes these local traces. Correct traces must then respect a well-formedness predicate ensuring the consistency of the composition without referring to program syntax or intermediate structures. Together, well-formed traces over states and events avoid complex data structures in configurations of reduction rules as it is the case in conventional SOS rules.

Compared to SOS, LAGC configurations contain not only the current state, but the whole trace leading to it, including any events that occurred. This richer structure makes it easy to extract information locally.

The modular separation of progress and composition is also crucial for our presentation. First, it permits an incremental presentation of LAGC for the different languages discussed here, introducing new concepts one at a time. Second, the fairness proof of our cooperative scheduler is completely independent of the syntax of sequential statements and only depends on events related to scheduling and synchronization.

## 4.1 States, Events, Traces, Continuations

For space reasons we refrain from giving full technical definitions of LAGC, but introduce essential notions in a compact manner. A fully precise account is in [12].

In SOS states map variables to concrete values. To permit *symbolic* expressions (containing variables) occurring as *values* in states, we use the star expression $*$ to represent an unknown value that cannot be further evaluated. The $*$ symbol does not occur in programs, only in the semantics. We adopt the notational convention of using capital letters for symbolic variables; but symbolic and non-symbolic variables belong to the same syntactic category and some operations transform a symbolic variable into a non-symbolic one.

**Definition 6** (Symbolic State, State Update)**.** *A symbolic state $\sigma$ is a partial mapping $\sigma : Var \rightharpoonup Sexp$ from variables to* symbolic expressions *$Sexp = Exp \cup \{*\}$. A symbolic variable is a variable $X$ bound to an unknown value $\sigma(X) = *$. Sexp are expressions that contain symbolic variables. The notation $\sigma[x \mapsto se]$ expresses the* update *of state $\sigma$ at $x$ with symbolic expression $se$.*

In a symbolic state $\sigma$, its symbolic variables symb($\sigma$) act as parameters, relative to which a local computation is evaluated. They are used to represent, for example, call parameters or task identifiers that cannot be known locally. We assume there are no dangling references[2] States without symbolic variables are called *concrete*. We denote with $\sigma \subseteq \sigma'$ that state $\sigma'$ extends (as a mapping) state $\sigma$.

**Example 3.** $\sigma = [x \mapsto Y + 42, Y \mapsto *]$ *is a symbolic state with* symb($\sigma$) = $\{Y\}$.

We assume an evaluation function $\mathrm{val}_\sigma : Sexp \to Sexp$ for symbolic expressions $se$ in the context of a state $\sigma$, defined as usual for concrete expressions, and as $\mathrm{val}_\sigma(X) = X$ for symbolic variables $X \in$ symb($\sigma$). It is always possible to evaluate expressions without symbolic variables to values and one could define a set of simplification rules on symbolic expressions, but they are not needed in the context of this article. The evaluation function is trivially extended to sets of expressions.

*Traces* are sequences over states and structured events. For example, the presence of synchronization events makes it possible to express global communication properties of traces via well-formedness conditions. Symbolic states imply symbolic traces, which motivates to constrain traces by path conditions.

**Definition 7** (Path Condition)**.** *A path condition $pc$ is a finite set of Boolean expressions. A fully evaluated concrete $pc$ is exactly one of $\emptyset$, $\{\mathtt{ff}\}$, $\{\mathtt{tt}\}$, $\{\mathtt{ff}, \mathtt{tt}\}$. It is* consistent *when it does not contain $\mathtt{ff}$. For any concrete state $\sigma$, path condition $\mathrm{val}_\sigma(pc)$ is fully evaluated.*

---

[2] A dangling reference is a reference to a variable not in the symbolic store.

**Definition 8** (Event Marker, Conditioned Symbolic Trace)**.**

*An* event marker *over expressions $\bar{e}$ is a term of the form $ev(\bar{e})$ .*

*A symbolic trace $\tau$ is defined inductively by the following rules ($\varepsilon$ denotes the empty trace):* $\qquad \tau ::= \varepsilon \mid \tau \frown t \qquad\qquad t ::= \sigma \mid ev(\bar{e}) \quad$.

*A conditioned symbolic trace has the form $pc \triangleright \tau$, where $pc$ is a path condition and $\tau$ is a symbolic trace. If $pc$ is consistent, we simply write $\tau$ for $pc \triangleright \tau$.*

Traces can be finite or infinite. Let $\langle \sigma \rangle$ denote the *singleton trace* $\varepsilon \frown \sigma$. *Concatenation* of two traces $\tau_1$, $\tau_2$ is written as $\tau_1 \cdot \tau_2$ and defined when $\tau_1$ is finite. The final state of a non-empty, finite trace $\tau$ is selected with $\mathrm{last}(\tau)$, the first state with $\mathrm{first}(\tau)$, respectively.

**Example 4.** *A conditioned symbolic trace is $\tau = \{Y > 0\} \triangleright \langle \sigma \rangle \frown \sigma[w \mapsto 17]$, where $\sigma$ is as in Example 3.*

Traces semantically model sequential composition of program statements. Assume $\tau_r$, $\tau_s$ are traces of statements $r$, $s$, respectively. To obtain the trace corresponding to sequential composition $r;s$, traces $\tau_r$ and $\tau_s$ must be concatenated, but the first state of the second trace should be identical to (or more precisely an extension of) the final state of the first trace. The chop operator gets rid of the redundant intermediate state.

**Definition 9** (Chop on Traces [16, 25])**.** *Let $pc_1, pc_2$ be path conditions and $\tau_1$, $\tau_2$ be symbolic traces, and assume that $\tau_1$ is a non-empty, finite trace. The* semantic chop $(pc_1 \triangleright \tau_1) ** (pc_2 \triangleright \tau_2)$ *is defined as follows:*

$$(pc_1 \triangleright \tau_1) ** (pc_2 \triangleright \tau_2) = (pc_1 \cup pc_2) \triangleright \tau \cdot \tau_2 \text{ where } \tau_1 = \tau \frown \sigma, \ \tau_2 = \langle \sigma' \rangle \cdot \tau' \text{ and } \sigma \subseteq \sigma' \ .$$

Chop is well-defined when the first argument is a finite non-empty trace, we only use it this way.

*Events* are uniquely associated with that state in a trace, where they occur. Events do not update a state, but may extend it with new symbolic variables. To do so, an event $ev(\bar{e})$ is inserted into a trace after a state $\sigma$, which can be extended by fresh symbolic variables $\bar{V}$, using an *event trace* $ev_\sigma^{\bar{V}}(\bar{e})$ of length three:

$$ev_\sigma^{\bar{V}}(\bar{e}) = \langle \sigma \rangle \frown ev(\mathrm{val}_{\sigma'}(\bar{e})) \frown \sigma', \text{ where } \sigma' = \sigma[\bar{V} \mapsto *].$$

Given a trace of the form $\tau_1 \frown \sigma$ and event $ev(\bar{e})$ with fresh symbolic variables $\bar{V}$, appending the event is achieved by the trace $\tau_1 \cdot ev_\sigma^{\bar{V}}(\bar{e})$. Def. 9 ensures that events in traces are joinable: $\tau ** ev_\sigma^{\bar{V}}(e)$ is well-defined whenever $\mathrm{last}(\tau) = \sigma$. If $\bar{V}$ is empty then the state is unchanged, in this case we omit the set of symbolic variables: $ev_\sigma(\bar{e}) = ev_\sigma^{\emptyset}(\bar{e})$.

**Example 5.** *To insert event $ev(Z)$, introducing symbolic variable $Z$, into trace $\tau$ from Example 4 at $\sigma$ we use the event trace $ev_\sigma^{\{Z\}}(Z) = \langle \sigma \rangle \frown ev(Z) \frown \sigma[Z \mapsto *]$. The resulting trace is: $\{Y > 0\} \triangleright \langle \sigma \rangle \frown ev(Z) \frown \sigma[Z \mapsto *] \frown \sigma[Z \mapsto *, w \mapsto 17]$.*

Traces are assumed to be well-formed, for example the domains of their states match, variables in events are defined, and so on [12].

Traces with symbolic variables model program executions relative to an unknown context. The symbolic variables in such traces become instantiated when the execution they represent is scheduled in a concrete context. At this point a symbolic trace is *concretised* by instantiating all of its symbolic variables. This results in a concrete trace with a path condition that is either consistent or not. Technically, we use the notion of a *concretisation mapping*. A concretisation mapping is defined relative to a state. It associates a concrete value to each symbolic variable of the state.

**Definition 10** (State Concretisation Mapping). *A mapping $\rho : Var \rightarrow Val$ is a concretisation mapping for a state $\sigma$ if $\text{dom}(\rho) \cap \text{dom}(\sigma) = \text{symb}(\sigma)$.*

A concretisation mapping $\rho$ may also define the value of variables not in the domain of $\sigma$. Concretisation mappings are canonically extended to events and conditioned traces [12].

**Example 6.** *Consider $\sigma$ of Example 3 with $\text{symb}(\sigma) = \{Y\}$. We define a concretisation mapping $\rho = [Y \mapsto 3]$ for $\sigma$ with $\rho(\sigma) = [x \mapsto 45, Y \mapsto 3]$. Applying $\rho$ to the trace in Example 4, we obtain $\rho(\tau) = \{3 > 0\} \triangleright \langle \rho(\sigma) \rangle \curvearrowright \rho(\sigma)[w \mapsto 17]$. We adopt the convention to strip away consistent path conditions such as here.*

The LAGC semantics evaluates one single statement "locally". Obviously, it is not possible to fully evaluate composite statements in this manner. Therefore, local LAGC rules perform one evaluation step at a time and defer evaluation of the remaining statements, which are put into a *continuation*, to subsequent rule applications. Syntactically, continuations are simply statements $s$ wrapped in the symbol $K$. To achieve uniform definitions we permit the case that no further evaluation is required (it has been completed) and use the "empty bottle" symbol for this.

**Definition 11** (Continuation Marker). *Let $s$ be a program statement or the symbol $\Diamond$. Then a* continuation marker *has the form $K(s)$.*

Local evaluation in Section 4.2 is defined such that for each statement $s$ and symbolic state $\sigma$ the result of $\text{val}_\sigma(s)$ is a set of conditioned, symbolic traces, so-called *continuation traces*, of the form $pc \triangleright \tau \cdot K(s')$ where $\tau$ is a *finite* trace. We denote by **CTr** the type of such traces. Let $\Theta$ be the set of traces of $s'$ and $\rho$ any concretisation mapping; then the expression $pc \triangleright \tau \cdot K(s')$ is used to describe the set of traces: $\{\rho(\tau) ** \tau' \mid \rho(pc) \text{ consistent}, \tau' \in \Theta\}$.

## 4.2 LAGC Semantics of While

We now define the LAGC semantics of While. Local evaluation rules accept a single statement in the context of a symbolic state $\sigma$ and return a set of finite continuation traces. We overload the symbol $\text{val}_\sigma$ with the type $\text{val}_\sigma : Stmt \rightarrow 2^{\textbf{CTr}}$ so that statements are also equipped with an evaluation function.

The rule for skip generates an empty path condition, returns the state it was called in, and produces the empty continuation. The result is one singleton trace:

$$\text{val}_\sigma(\text{skip}) = \{\emptyset \triangleright \langle \sigma \rangle \cdot K(\Diamond)\} \ .$$

The assignment rule generates an empty path condition and a trace from the current state $\sigma$ to a state which updates $\sigma$ at $x$, and produces the empty continuation.

$$\mathrm{val}_\sigma(x := e) = \{\emptyset \triangleright \langle \sigma \rangle \frown \sigma[x \mapsto \mathrm{val}_\sigma(e)] \cdot \mathrm{K}(\lozenge)\} \ .$$

The conditional statement is a complex statement and cannot be evaluated locally in one step, so we expect it to produce a non-empty continuation. The rule branches on the value of the condition, resulting in two traces with complementary path conditions. The first trace is obtained from the current state and the continuation corresponding to the if-branch, and the second trace corresponds to the empty else-branch:

$$\mathrm{val}_\sigma(\text{if } e \ \{ \ s \ \}) = \{\{\mathrm{val}_\sigma(e)\} \triangleright \langle \sigma \rangle \cdot \mathrm{K}(s), \ \ \{\mathrm{val}_\sigma(!e)\} \triangleright \langle \sigma \rangle \cdot \mathrm{K}(\lozenge)\} \ . \tag{1}$$

The rule for sequential composition $r;s$ is obtained by first evaluating $r$ to traces of the form $pc \triangleright \tau \cdot \mathrm{K}(r')$ with continuation $r'$. Then $s$ is added to this continuation:

$$\mathrm{val}_\sigma(r;s) = \{pc \triangleright \tau \cdot \mathrm{K}(r';s) \mid pc \triangleright \tau \cdot \mathrm{K}(r') \in \mathrm{val}_\sigma(r)\} \ .$$

If $r'$ happens to be the empty continuation $\lozenge$ it must be ignored. To achieve this the rewrite rule "$\lozenge; s \rightsquigarrow s$" is exhaustively applied to statements inside continuations. We use the semantics of while to illustrate that the semantics of a statement can be expressed in terms of the semantics of other statements (here if and sequence), without having to expose intermediate states:

$$\mathrm{val}_\sigma(\text{while } e \ \{ \ s \ \}) = \mathrm{val}_\sigma(\text{if } e \ \{ \ s; \ \text{while } e \ \{ \ s \ \}\}) \ .$$

Note that this definition is not circular because the evaluation of if puts the while statement inside a continuation.

**Example 7.** *We start evaluation of the statement $s_{seq} = (x := 1; \ y := x + 1)$ in an arbitrary symbolic state $\sigma$. The rule for sequential composition yields $\mathrm{val}_\sigma(s_{seq}) = \{\emptyset \triangleright \langle \sigma \rangle \frown \sigma[x \mapsto 1] \cdot \mathrm{K}(y := x + 1)\}$ . It uses the result of evaluating the first assignment in the context of $\sigma$: $\mathrm{val}_\sigma(x := 1) = \{\emptyset \triangleright \langle \sigma \rangle \frown \sigma[x \mapsto 1] \cdot \mathrm{K}(\lozenge)\}$ .*

In the composition rules below it might happen that the empty continuation is evaluated, so there must be a rule for it. Local evaluation returns a set of traces and the empty continuation indicates that nothing more is to be evaluated, hence its evaluation yields the empty set of traces: $\mathrm{val}_\sigma(\lozenge) = \{\}$.

Local traces are instantiated and composed into concrete global ones. As While is sequential and deterministic, there is exactly one trace, provided that the execution starts in a concrete state that assigns values to all the variables of a program [12]. In consequence, no scheduler needs to be defined for While.

The task of the composition rule for While programs is to evaluate one statement at a time in a concrete state until the next continuation, then stitch the resulting concrete traces together. Given a configuration with *concrete* trace $sh$ having final state $\sigma$ and a continuation $\mathrm{K}(s)$, we evaluate $s$ starting in $\sigma$. The result is a set[3] of conditioned traces from which one trace with a consistent path condition and a trailing continuation $K(s')$ is chosen. The chosen trace $\tau$ is joined with the given trace $sh$. Afterwards, the composition rule can be applied again to the extended concrete trace and $K(s')$.

---

[3] A singleton set in the case of While, but the same rule will be used later for non-deterministic languages.

$$\frac{\sigma = \text{last}(sh) \quad pc \triangleright \tau \cdot K(s') \in \text{val}_\sigma(s) \quad pc \text{ consistent}}{sh, K(s) \rightarrow sh \ast\ast \tau, K(s')} \tag{2}$$

The rule assumes that $s$ is evaluated to a concrete trace $\tau$ so $sh \ast\ast \tau$ stays concrete. At this stage, symbolic traces cannot occur. The global semantics of While programs results from the transitive closure of the transition defined by rule (2).

## 5  Scheduling a Language with Spawn

This section presents our approach in a simple setting. We introduce concurrency via the Spawn language in Figure 2 that spawns threads when calling a procedure, and implements an interleaving semantics. A simple fair scheduler executes each task in a round-robin manner. We specify and prove fairness for that case.

### 5.1  LAGC Semantics

We provide a semantics of Spawn based on the principles of LAGC: first a local semantics is given that merely emits a spawn event whenever a thread is spawned, and it uses a spawn reaction event at the points where a thread can be actually created (here, at the beginning of the execution of a procedure). When composing the semantics of different threads, we use a *well-formedness* predicate to ensure those events match.

For a state $\sigma$, the semantics of Spawn has events $callEv(m, \text{val}_\sigma(e))$, $callREv(m, \text{val}_\sigma(e))$. These events denote the *call* and the *activation* (i.e., *call reaction*) of a procedure $m$ with argument $e$, respectively. Recall from Section 4.1 that events in a trace are preceded and succeeded by associated states. The local evaluation rule (3) for a call to $m$ with argument $e$ has an empty path condition, inserts a *call* event $callEv_\sigma(m, \text{val}_\sigma(e))$ into the trace at $\sigma$, and produces the empty continuation. This makes the call *non-blocking*: the code following the call can be executed immediately, if scheduled.

$$\text{val}_\sigma(\text{spawn}(m, e)) = \{\emptyset \triangleright callEv_\sigma(m, \text{val}_\sigma(e)) \cdot K(\llbracket\rrbracket)\} \ . \tag{3}$$

The composition rules for Spawn combine locally evaluated traces into global ones by exploring the multiset of continuations for each thread. In this simple setting, procedure calls and processors are anonymous in the sense that they lack task identifiers. Consequently, it is possible that two tasks are identical. For this reason, we represent continuation candidates as a multiset $p$, denoted with "⊎" the *disjoint* union of multisets (and sets). We add a rewrite rule to simplify empty continuations in the multisets of tasks—this rule is applied exhaustively when adding tasks in multisets: $p \uplus \{K(\llbracket\rrbracket)\} \rightsquigarrow p$. The composition rules extend partial traces by one step and have the form: $sh, p \rightarrow sh', p'$ where $p, p'$ are multisets of continuations of the form $K(s)$.

First we discuss the rule corresponding to rule (2). The difference between rule (2) and rule (4) is that we no longer commit to a single continuation, because several procedure bodies can be evaluated. The rule selects and removes *one* matching continuation from the pool $p \uplus \{K(s)\}$, executes it until the next continuation marker, exactly like rule (2), and then puts the code remaining to be executed back into $p$.

$$\frac{\sigma = \text{last}(sh) \qquad pc \rhd \tau \cdot K(s') \in \text{val}_\sigma(s) \qquad pc \text{ consistent}}{sh, p \uplus \{K(s)\} \to sh \ast\ast \tau, p \uplus \{K(s')\}} \tag{4}$$

We need a second composition rule that adds procedure bodies to the pool, and creates a new execution thread. We select a procedure with body $s$ from table $\overline{M}$ and create a new continuation $K(s[x \leftarrow y])$ corresponding to its body, where the call parameter $x$ is substituted with a fresh variable $y$ for disambiguation and a corresponding entry is added to the state $\sigma$. This continuation is added to the pool $p$.

$$\frac{m(x)\{s\} \in \overline{M} \qquad \sigma = \text{last}(sh) \qquad wf(sh \ast\ast callREv_\sigma(m, v)) \qquad y \notin \text{dom}(\sigma)}{sh, p \to sh \ast\ast callREv_\sigma(m, v) \frown \sigma[y \mapsto v], p \uplus \{K(s[x \leftarrow y])\}} \tag{5}$$

We must take care that rule (5) is only triggered when a corresponding sᴘᴀᴡɴ happened in $sh$. This is ensured by combining two mechanisms: events and well-formedness predicates. Recall that local rule (3) for sᴘᴀᴡɴ emits a call event with the procedure name $m$ and an evaluated procedure parameter $v$. We keep track of which calls have been activated using these events. The third premise of rule (5) guesses an activation event $callREv_\sigma(m, v)$ whose call event in $sh$ had not been resolved yet. The obtained matching yields procedure $m$ that is added to the pool $p$, as well as value $v$ that is assigned to the formal parameter $y$ (since $y$ is fresh, it can only appear in the continuation). Procedures are anonymous and only distinguished by name and argument. In consequence, well-formedness (Def. 12) boils down to counting the number of unresolved calls with the same name and argument. Rule (4) cannot violate well-formedness and rule (5) ensures it, hence, well-formedness is an invariant of the generated traces.

**Definition 12** (Well-Formedness)**.** *Whenever a trace $sh$ is extended with a call reaction event $callREv(m, v)$, there must be a corresponding call event $callEv(m, v)$ in $sh$. This is ensured by counting the number of occurrences of both events in $sh$ with $\#_{sh}(ev(\overline{\varepsilon}))$. In all other cases, the trace stays well-formed when it is extended with a new element. $wf(sh)$ is defined inductively over $sh$:*

$$wf(\varepsilon) = true$$
$$wf(sh \frown \sigma) = wf(sh) \qquad\qquad wf(sh \frown callEv(m, v)) = wf(sh)$$
$$wf(sh \frown callREv(m, v)) = wf(sh) \wedge \#_{sh}(callEv(m, v)) > \#_{sh}(callREv(m, v))$$

An execution of a sᴘᴀᴡɴ program with main block $s_{main}$ starts from an initial configuration $\langle \sigma_\epsilon, \{K(s_{main})\}\rangle$; where $\sigma_\epsilon$ is initializes variables of the program to 0.

## 5.2 A Concrete Weakly Fair Scheduler for Spawn

As in Section 3.3.3, we define a notion of fairness aligned with language semantics in the sense that it is based on reachable statements. We adapt Def. 4 to the case of LAGC, where statements are given by continuations in the configuration. Configurations in LAGC are pairs of traces and multisets of continuations. Only the latter are needed to define reachable statements: $\text{newStmt}((sh, p), (sh', p')) = p' - p$. Then $\text{enabled}(sh, p)$, weak and strong fairness are defined exactly as in Section 3.3.3 (Defs. 4, 5), except that statements $s$ are replaced with continuations $K(s)$.

To define a concrete fair scheduler we replace the multiset representation of continuations in the trace composition rules with a circular buffer. For our simple procedure calls, this buffer just contains continuations. The buffer is implemented as a *queue q* over continuations of the form $K(s)$. We can now reformulate rules (4)–(5) as follows:

$$\frac{\sigma = \text{last}(sh) \quad \text{pop}(q) = (q', K(s)) \quad pc \triangleright \tau \cdot K(s') \in \text{val}_\sigma(s) \quad pc \text{ consistent}}{\neg wf(sh \ast\ast callREv_\sigma(m, v)) \text{ for any } m(x)\{s\} \in \overline{M}}{sh, q \to_S sh \ast\ast \tau, \text{push}(q', K(s'))} \quad (6)$$

$$\frac{m(x)\{s\} \in \overline{M} \quad \sigma = \text{last}(sh) \quad wf(sh \ast\ast callREv_\sigma(m, v)) \quad y \notin \text{dom}(\sigma)}{sh, q \to_S sh \ast\ast callREv_\sigma(m, v) \frown \sigma[y \mapsto v], \text{push}(q, K(s[x \leftarrow y]))} \quad (7)$$

Rule (7) eagerly starts execution of a procedure body, whenever possible, i.e. whenever there is an unresolved matching call event. The code in the procedure body is pushed to the end of the queue. Only if rule (7) is not applicable then rule (6) can be applied. This is ensured by the premise on the second line. The rule executes a statement of the first continuation $K(s)$ in $q$ and pushes the remaining $K(s')$ to its rear.

It is possible that neither rule (6) nor (7) is applicable, but there are still continuations remaining to be executed. Such a situation arises when the first element of $q$ is $K(\emptyset)$. In this case, $\text{val}_\sigma(\emptyset) = \{\}$ and rule (6) is inapplicable. If, at the same time, there are no unresolved procedure calls, rule (7) cannot be applied either. The following rule advances to the next queue element and disposes of the empty continuation:

$$\frac{\text{pop}(q) = (q', K(\emptyset)) \quad \neg wf(sh \ast\ast callREv_\sigma(m, v)) \text{ for any } m(x)\{s\} \in \overline{M}}{sh, q \to_S sh, q'} \quad (8)$$

The initial judgment for producing traces of a program $P$ with a main body $s_{main}$ has the form "$\langle \sigma_e \rangle, q_{ini}$", where $q_{ini} = \text{push}(\text{create}(), K(s_{main}))$.

It is possible that a transition of the LAGC scheduler semantics has no counterpart in the original LAGC semantics. This is the case for rule (8) and will get more pronounced in the next section, where some tasks might not be able to progress.

The LAGC semantics of While is deterministic up to naming of fresh variables. This is not the case for the Spawn language, as seen in rules (4)–(5), because of multi-threading. However, the scheduler, defined by rules (6)–(8) is deterministic, so these rules are a suitable basis for a scheduler implementation. The following proposition states the determinism property.

**Proposition 1.** *Starting with an initial judgment of the form "⟨σε⟩, qini", consider a well-formed trace sh and non-empty queue q reachable by application of rules (6)–(8). From sh, q exactly one of the rules (6)–(8) is applicable and the result of the rule application is deterministic up to renaming of variables. In other words, the scheduler is deterministic.*

The proof is in App. B. The main result of the present section is weak fairness of the deterministic scheduler, as stated in the following theorem, where $M(q)$ is the *multiset* of non-empty continuations inside a queue $q$:

**Theorem 1.** *Starting from an initial judgment, the scheduler defined by rules (6)–(8) is one possible weakly fair scheduler of our semantics: If ⟨σε⟩, qini →S sh1, q1 →S sh2, q2 →S ⋯ then there is a weakly fair execution ⟨σε⟩, {K(smain)} →\* sh1, M(q1) →\* sh2, M(q2) →\* ⋯ that produces the same trace.*

App. C proves this theorem. The proof is based on the definition of an adequate *scheduling distance* of a given statement that is positive and decreases for a continuation belonging to enabled($sh, p$). There is no blocking statement, and thus every thread admits one continuation for which the path condition is consistent. Every thread that can be enabled will stay enabled until scheduled. Consequently, there is no difference between strong and weak fairness and the theorem holds for *strong fairness*. The next section introduces guarded statements that differentiate between weak and strong fairness, and require additional queue manipulations to find an executable task.

## 6 Adding Guarded Commands

**LAGC Semantics**  We locally evaluate a `Guard` statement (see Section 2) such that it only progresses when the guard is true.[4] All other local evaluation rules are unchanged.

$$\mathrm{val}_\sigma(:: g; s) = \{\{\mathrm{val}_\sigma(g)\} \triangleright \langle \sigma \rangle \cdot K(s)\}\} \tag{9}$$

This rule looks innocent, but has considerable consequences: the only local evaluation rule introducing path conditions so far is rule (1). It produces traces with complementary path conditions, so rule (4) is *always* applicable to any $K(s) \in q$ with $s \neq \emptyset$. In case of rule (9) and $s = (:: g; s')$ the sole path condition might be inconsistent. In particular, it is possible that none of the scheduling rules defined in Section 5.2 is applicable. More precisely, the scheduler defined there can get stuck when $K(:: g; s') \in \mathrm{pop}(q)$ and the path condition is inconsistent.

**Scheduling**  Our scheduler skips a stuck continuation and puts it at the end of the queue. To do so, we use a helper function rotate($q$) that moves the first element of the queue to its rear: $\text{rotate}(q) = \text{push}(q', k)$, where $(q', k) = \text{pop}(q)$. We use this operation to rotate the queue when none of the previous scheduling rules are applicable:

$$\frac{\text{pop}(q) = (q', \text{K}(s)) \quad \neg wf(sh ** callREv_\sigma(m, v)) \text{ for any } m(x)\{s\} \in \overline{M}}{\left(\nexists\, pc, \tau, s'.\, pc \triangleright \tau \cdot \text{K}(s') \in \text{val}_\sigma(s) \wedge pc \text{ consistent}\right) \quad s \neq \emptyset}{sh, q \rightarrow_S sh, \text{rotate}(q)} \tag{10}$$

---

[4] Different (LAGC) semantics can be chosen for this statement. In particular, a semantics with an inconsistent path that "stutters" is discussed in [12].

**Proving Fairness**  At first glance, rule (10) looks like bad news for fairness, because rotating an unusable continuation to the end increases its distance to a scheduling point, possibly preventing the guard statement to become ever scheduled. Consider the following example.

**Example 8.** *Consider a scheduling queue containing the following four statements in the given sequence:*

| *while true do { b = true }* | *:: !b; s* | *while true do { b = false }* | *:: b; s* |

*The guarded statements are never scheduled, because the guard is set to the complement by the preceding assignment in the loop body.*

This example illustrates that our scheduler indeed does not achieve strong fairness. However, it achieves weak fairness because a continuation that qualifies for the weak fairness condition is always schedulable and thus not rotated to the end of the task queue. The example also hints that to properly reason on rotation it is necessary to distinguish carefully between continuation statements that coincide syntactically, but derive from different code contexts (such as s). Theorem 1 still holds, but the proof needs to be adapted, it is detailed in App. D.

## 7  Cooperative Scheduling of a Language with Spawn, Suspend, and Join

This section discusses the `CoopWhile` language (see Section 2). As for `Spawn` and `Guard`, its semantics is an extension of the previous sections, benefiting from the modularity of the LAGC approach. Then we define a scheduler that provenly ensures quiescent fairness for `CoopWhile`.

### 7.1  LAGC Semantics of `CoopWhile`

In the semantics of `CoopWhile`, the evaluation function must be aware of the task identifier for the current task. This task identifier cannot possibly be known locally. Hence, the form of the evaluation function becomes $\text{val}_\sigma^F(s)$, where $F$ is a symbolic task identifier typed with *TId*. To keep track of task identifiers in configurations, we associate a *concrete* task identifier $f$ with each continuation in the composition rules, so these take now the form $\text{K}^f(s)$.

It was shown in [14] that the semantics of cooperative scheduling can be characterized with the help of four event types: *callEv*$(e, m, f)$ ("call") records spawning of task $m$ with argument $e$ and task identifier $f$; *callREv*$(v, m, f)$ ("call reaction") records that task $m$ with argument value $v$ and task identifier $f$ started to execute, while $compEv(f)$ ("completion") signals that it finished. Finally, $compREv(f)$ ("completion reaction") records the reaction of a statement waiting for the task identified by $f$ to finish. While the first three event types are unique for $f$ in any trace, there could be several (or no) completion reaction events.

The local evaluation rule for spawning a task just emits a call event. The task identifier $F'$ of the spawned task needs to be recorded in it and serves to identify the spawned task. $F'$ is a fresh symbolic variable since its value cannot be known locally.

$$\mathrm{val}_\sigma^F(\mathsf{spawn}(m,e,x)) = $$
$$\{\emptyset \triangleright callEv_\sigma^{\{F'\}}(\mathrm{val}_\sigma^F(e),m,F') \frown \sigma[x \mapsto F', F' \mapsto *] \cdot \mathrm{K}(\mathbb{0}) \mid F' \notin \mathrm{dom}(\sigma)\} \qquad (11)$$

The rule for evaluating a procedure body emits a call reaction event with a task identifier that matches the current task identifier $F$. Well-formedness in the composition rules ensures that the call reaction event is matched to a preceding call event that has the same value of $F$. To mark the syntactic end of a procedure body, which signifies its completion, a return statement is appended to the code in the continuation. Like in the SOS semantics, this return statement is only part of the *internal* syntax introduced by the semantics does not appear in the original program.

$$\mathrm{val}_\sigma^F(m(x)\ \{s\}) = $$
$$\{\emptyset \triangleright callREv_\sigma^{\{X\}}(X,m,F) \frown \sigma[x' \mapsto X, X \mapsto *] \cdot \mathrm{K}(s[x \leftarrow x'];\ \mathsf{return}) \mid x', X \notin \mathrm{dom}(\sigma)\}$$
$$(12)$$

The presence of return now triggers a completion event for the current task identifier:

$$\mathrm{val}_\sigma^F(\mathsf{return}) = \{\emptyset \triangleright compEv_\sigma(F) \cdot \mathrm{K}(\mathbb{0})\} \qquad (13)$$

The semantics for suspending statements relies on the fact that any code starting with await or suspend can be suspended. Moreover, a suspended task is only re-scheduled when it can progress. This invariant is ensured by the trace composition rules below. In other words, it is the composition rules that deal with suspension when encountering the await. Local evaluation of await needs only specify how the task is re-activated. Consequently, the semantics of await merely needs to emit a completion reaction event. This event must match a previous completion event involving the same task identifier in its argument, which is ensured by well-formedness. Similarly, *suspend*, when re-scheduled, simply has the semantics of skip.

$$\mathrm{val}_\sigma^F(\mathsf{await}\,x) \quad = \quad \{\emptyset \triangleright compREv_\sigma(\mathrm{val}_\sigma^F(x)) \cdot \mathrm{K}(\mathbb{0})\} \qquad (14)$$
$$\mathrm{val}_\sigma^F(\mathsf{suspend}) \quad = \quad \{\emptyset \triangleright \langle \sigma \rangle \cdot \mathrm{K}(\mathbb{0})\} \qquad (15)$$

Since there is only one processor, there is at most one active task, all other tasks are suspended, a task is suspended if it is the continuation of a suspended statement:

**Definition 13** (Suspended Statement, Active Task)**.** *A statement of the form* "await $x$; $s$" *or* "suspend; $s$" *is called a* suspended statement *and can be tested with* suspended($s$). *The symbol $Q$ denotes a set of* suspended *statements, i.e. a (possibly empty) set of* *continuations of the form $\mathrm{K}^f(s)$ such that $s$ is either a suspended statement or the empty continuation $\square$.*

*The current task queue of a program trace is represented by a set of continuations that is either empty or of the form $Q \uplus \{\mathrm{K}^f(s)\}$. If $s$ is a suspended statement then no task is currently executing and any schedulable task can be activated. Otherwise, $\mathrm{K}^f(s)$ is the* only *continuation that can be executed and is called the* active *task.*

Note that there is at most one active task in a valid configuration. Any statement that is not suspended and that has a consistent path condition can be executed. We assume the rewrite rule "$Q \uplus \{\mathrm{K}^f(\square)\} \rightsquigarrow Q$", discarding empty continuations, is applied to continuation sets whenever possible. Hence, we can assume $Q$ consists only of suspended statements. We define two trace composition rules. Recall that local evaluation rules (11)–(12) contain symbolic variables, because the caller id and value cannot be known locally. These variables need to be instantiated during application of the trace composition rules when a procedure is spawned and executed by a suitable concretisation mapping $\rho$ that preserves well-formedness of the generated trace.

Rule (16) handles the case when a task progresses. Configurations have the form $sh, Q \uplus \{\mathrm{K}^f(s)\}$. If there is an active task this must be $s$, it is executed until it completes or suspends. Otherwise, $s$ is a suspended statement and rules (14)–(15) apply.

$$\frac{\sigma = \mathrm{last}(sh) \qquad pc \triangleright \tau \cdot \mathrm{K}(s') \in \mathrm{val}_\sigma^f(s)}{sh, Q \uplus \{\mathrm{K}^f(s)\} \rightarrow sh \ast\ast \rho(\tau), Q \uplus \{\mathrm{K}^f(s')\}} \tag{16}$$

$$\frac{\rho \text{ concretizes } \tau \qquad \rho(pc) \text{ consistent} \qquad wf(sh \ast\ast \rho(\tau))}{sh, Q \uplus \{\mathrm{K}^f(s)\} \rightarrow sh \ast\ast \rho(\tau), Q \uplus \{\mathrm{K}^f(s')\}} \tag{16}$$

When there is no active task (only $Q$ is present in the starting configuration of the conclusion), rule (17) is applicable as well: it starts the execution of a new task that matches an unresolved call event $callEv(v, m, f)$ in $sh$. The evaluation function is tagged with the concrete task identifier $f$, whose freshness is ensured by well-formedness. As no task is active, the spawned task can start running immediately.

$$\frac{m(x)\{s\} \in \overline{M} \qquad \sigma = \mathrm{last}(sh) \qquad f \in TId \qquad pc \triangleright \tau \cdot \mathrm{K}(s') \in \mathrm{val}_\sigma^f(m(x)\,\{s\})}{sh, Q \rightarrow sh \ast\ast \rho(\tau), Q \uplus \{\mathrm{K}^f(s')\}} \tag{17}$$

$$\frac{\rho \text{ concretizes } \tau \qquad \rho(pc) \text{ consistent} \qquad wf(sh \ast\ast \rho(\tau))}{sh, Q \rightarrow sh \ast\ast \rho(\tau), Q \uplus \{\mathrm{K}^f(s')\}} \tag{17}$$

Well-formedness rules ensure the following invariants hold: (i) a unique task identifier is associated with each task (first line), (ii) each task starts only after it was spawned (second line), and (iii) an await statement is only re-activated once the designated task is completed (fourth line). The well-formedness rules are defined inductively on the length of traces by case distinction on the event type to be appended to an existing trace $sh$.

$$wf(sh \frown callEv(v, m, f)) = wf(sh) \wedge \nexists v', m'. callEv(v', m', f) \in sh$$
$$wf(sh \frown callREv(v, m, f)) = wf(sh) \wedge callEv(v, m, f) \in sh \wedge callREv(v, m, f) \notin sh$$
$$wf(sh \frown compEv(f)) = wf(sh)$$
$$wf(sh \frown compREv(f)) = wf(sh) \wedge compEv(f) \in sh$$

No rule for $compEv(f)$ is required: By rule (13) $compEv(f)$ terminates a task that was initiated by a matching $callREv(v, m, f)$ event and thus corresponds to a well-defined spawn.

An execution of a program starts from an initial configuration $\langle \sigma_\epsilon, \{K^{f^{init}}(s_{main})\}\rangle$ and has the form: $sh_0, Q_0 \rightarrow^* sh_m, Q_m \uplus \{K^{f_m}(s_m)\} \rightarrow^* sh_n, Q_n \uplus \{K^{f_n}(s_n)\} \rightarrow \cdots$.

## 7.2 Fair Cooperative Scheduling of CoopWhile

The definition of reachable statements is unchanged: for LAGC semantics it is generic. We turn to refurbishing quiescent fairness (Def. 5) for the LAGC semantics, where enabled statements derive from the continuations in an LAGC configuration. We need one minor change: Def. 2 of quiescent states is based on the idle configurations in the SOS, in LAGC a *quiescent state* is a configuration of the form $sh, Q$. In addition, the notions of enabled statement and fairness are over multisets of the form $K^f(s)$, i.e., take into account the task identifier. [5]

The scheduling rules need to be explicit about which task is currently active (if any) and which tasks are waiting to be executed. Hence a scheduler configuration C is either of the form $(K^f(s), q)$ or $(Idle, q)$, where the first element is the active task (Idle, when there is no active task) and $q$ is a queue of elements in $\{K^f(await\ x; s), K^f(suspend; s), (m, f)\}$. Elements of $q$ are either suspended tasks or *task creation markers* of the form $(m, f)$.

Task creation must be performed only once for each task identifier and can be performed immediately after a call. Rule (18) creates a task marker $(m, f)$. It is triggered immediately once a call event involving $m$ and $f$ is added to $sh$. The $doneEv$ at the end of $sh$ prevents the rule from being triggered repeatedly. The task marker $(m, f)$ is pushed on the queue.

$$\frac{lastEvent(sh) = callEv(v, m, f)}{sh, (active, q) \rightarrow_S sh ** doneEv, (active, push(q, (m, f)))} \tag{18}$$

The remaining rules do the actual scheduling and are structured along the shape of the current scheduler configuration C, as depicted in Figure 6. We start on the left with the case when the processor is idle with a queue $q$ of waiting tasks. Its first element, obtained by $pop(q)$, determines which rule is taken. In case of $(m, f)$, starting a procedure is always possible and this is performed by rule (19):

$$\frac{\nexists\ v, m, f'. lastEvent(sh) = callEv(v, m, f') \qquad f \in TId \qquad \sigma = last(sh)}{sh, (Idle, q) \rightarrow_S sh ** \rho(\tau), (K^f(s'), q')} \\ \frac{pop(q) = (q', (m, f)) \qquad m(x)\{s\} \in \overline{M} \qquad pc \triangleright \tau \cdot K(s') \in val_\sigma^f(m(x)\{s\})}{\rho\ concretizes\ \tau \qquad \rho(pc)\ consistent \qquad wf(sh ** \rho(\tau))} \tag{19}$$

When $q$'s first element is a suspended statement $s$, then this may or may not be schedulable in $sh$. If it is schedulable then $s$ is evaluated as usual. By definition of $q$ the statement $s$ must be a suspended statement, so no explicit premise stating this fact is needed.

---

[5] Enabled statements could also be defined by forgetting the task identifier when defining the newStmt set, with similar results.

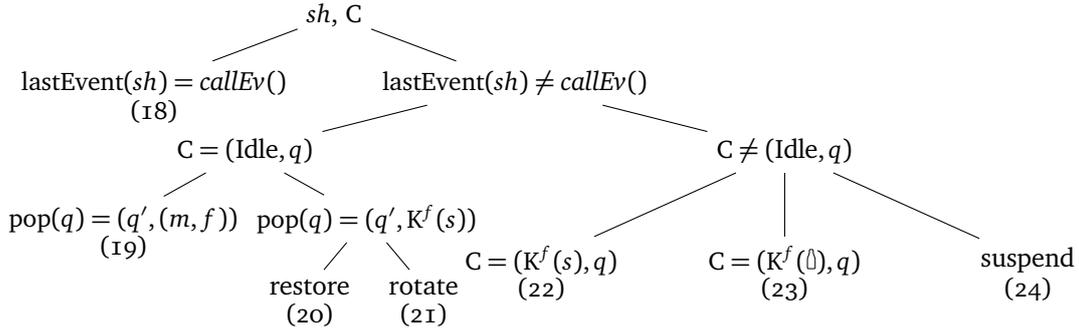

**Figure 6** Structure of the rules that schedule the spawn/suspend/join language

$$\frac{\begin{array}{cc} \nexists\, v, m, f'.\, \text{lastEvent}(sh) = callEv(v, m, f') & \sigma = \text{last}(sh) \\ \text{pop}(q) = (q', K^f(s)) \quad pc \triangleright \tau \cdot K(s') \in \text{val}_\sigma^f(s) \quad \rho \text{ concretizes } \tau \\ \rho(pc) \text{ consistent} \quad wf(sh \ast\ast \rho(\tau)) \end{array}}{sh, (\text{Idle}, q) \rightarrow_S sh \ast\ast \rho(\tau), (K^f(s'), q')} \quad (20)$$

When $q$'s first element is not schedulable, we rotate $q$ and the processor stays idle. This happens when the task waits for completion of another task that has not finished. The premise in the second line states that there is no way to find a well-formed trace that permits the first schedulable task to continue.

$$\frac{\begin{array}{cc} \nexists\, v, m, f'.\, \text{lastEvent}(sh) = callEv(v, m, f') & \text{pop}(q) = (q', K^f(s)) \\ \big(\exists\, x, s'.\, s = \text{await } x;\, s' \wedge \sigma = \text{last}(sh) \wedge compEv(\sigma(x)) \notin sh\big) \end{array}}{sh, (\text{Idle}, q) \rightarrow_S sh, (\text{Idle}, \text{rotate}(q))} \quad (21)$$

When $q$ is empty there is no remaining task to be executed, no scheduler rule is applicable, so final configurations have the form $sh, (\text{Idle}, \emptyset)$. In contrast, a configuration of the form $sh, (\text{Idle}, q)$, where $q \neq \emptyset$ and no rule is applicable, indicates a deadlock.

We are left with the cases when the active task is *not* idle, i.e. it is a continuation $K^f(s)$. If $s$ is not a suspended statement it can be evaluated. The resulting rule (22) is very similar to (16) with a different structure for configurations; it also has additional premises to make the scheduler deterministic:

$$\frac{\begin{array}{cc} \nexists\, v, m, f'.\, \text{lastEvent}(sh) = callEv(v, m, f') & \sigma = \text{last}(sh) \\ \neg\text{suspended}(s) \quad pc \triangleright \tau \cdot K(s') \in \text{val}_\sigma^f(s) \quad \rho \text{ concretizes } \tau \\ \rho(pc) \text{ consistent} \quad wf(sh \ast\ast \rho(\tau)) \end{array}}{sh, (K^f(s), q) \rightarrow_S sh \ast\ast \rho(\tau), (K^f(s'), q)} \quad (22)$$

If $s$ cannot continue this can be for two reasons: either $s = \Box$, i.e. the current task is completed, the processor becomes Idle, and the empty continuation is disposed of:

$$\frac{\nexists\, v, m, f.\, \text{lastEvent}(sh) = callEv(v, m, f')}{sh, (K^f(\Box), q) \rightarrow_S sh, (\text{Idle}, q)} \quad (23)$$

The last case is when $s$ is a suspended statement. In this case, the current task is suspended: it is pushed to the end of queue $q$:

$$\frac{\nexists\, v, m, f'.\, \text{lastEvent}(sh) = callEv(v, m, f') \qquad \text{suspended}(s)}{sh, (\text{K}^f(s), q) \rightarrow_S sh, (\text{Idle}, \text{push}(q, \text{K}^f(s)))} \qquad (24)$$

It is obvious that the preconditions of the scheduling rules are mutually exclusive and exhaustive. Moreover, it is easy to see that the rule set is deterministic, provided a single consistent continuation can be derived from each statement. This is the case in the CoopWhile language. To be precise, the only source of non-determinism in the rules is the choice of fresh names for variables and task identifiers. Consequently, rules (18)–(24) constitute a well-defined, deterministic scheduler.

To state fairness, we extend the translation M in Thm. 1 for the new scheduler configurations. This is straightforward: M collects the set of tasks that are neither Idle nor empty continuations and it skips task markers. Translation of traces simply removes *doneEv* events. With such a translation function M the following theorem holds, i.e. our scheduler ensures a quiescent fair execution. See App. E for details.

**Theorem 2** (Quiescent Fairness). *Starting from an initial judgment, the scheduler defined by rules (18)–(24) is a quiescent fair scheduler for CoopWhile: If $\langle \sigma_\varepsilon \rangle$, $\text{C}_S^{init} \rightarrow_S sh_1$, $\text{C}_1 \rightarrow_S sh_2$, $\text{C}_2 \rightarrow_S \cdots$ then there is a quiescent fair execution (LAGC semantic evaluation) $\langle \sigma_\varepsilon \rangle$, $\{\text{K}^{f^{init}}(main)\} \rightarrow^* \text{M}(sh_1)$, $\text{M}(\text{C}_1) \rightarrow^* \text{M}(sh_2)$, $\text{M}(\text{C}_2) \rightarrow^* \cdots$ that produces the same trace up to doneEv events.*

Again, the execution in the original LAGC semantics has less steps than the scheduler execution, because the scheduler rotates tasks and inspects the tasks that are unable to progress. The proof given in App. E relies on an extension of the scheduling distance of a given statement used previously. The scheduling distance is minimal when a statement is the active task and it decreases otherwise, provided that the statement still qualifies for being scheduled later. But if it does not qualify then the condition of quiescent fairness is violated for the statement.

## 8    Comparison of SOS and LAGC Semantics

We compare the SOS-style semantics of CoopWhile in Figure 4 with the LAGC-style semantics given in Section 7.1. One can note that LAGC is more *implicit* and *abstract*: In LAGC there is only a single progress rule (16) for any kind of statement and the active task identifier $f$ is implicitly given in a task queue of the shape $Q \uplus \{\text{K}^f(s)\}$. The applicability of each "local" evaluation rule (11)–(15) is governed by well-formedness.

In contrast, there is a different SOS rule for each kind of statement $s$. This permits some simplifications in the SOS rules. For example, since there is an explicit progress rule for spawn, it is possible to combine spawning with "starting" a procedure, i.e. to put its (renamed) body with a fresh task identifier in the task list. Consequently, no rule corresponding to rule (17) is needed in the SOS rules and no task markers $(m, f)$ in the scheduler version. Accordingly, the SOS scheduler in Figure 5 has less rules than the LAGC scheduler in Section 7.2.

$$f \cdot (:: g; s), \, T, \, \sigma \to \text{Idle}, \text{push}(q, (f, (:: g; s))), \, \sigma \quad \text{YieldGuard}$$

$$\frac{\text{pop}(q) = (q', (f, s)) \qquad s = (:: g; s') \qquad \text{val}_\sigma(g) = \text{True}}{\text{Idle}, \, q, \, \sigma \to f \cdot s', \, q', \, \sigma} \quad \text{ScheduleGuardDone}$$

$$\frac{\text{pop}(q) = (q', (f, s)) \qquad s = (:: g; s') \qquad \text{val}_\sigma(g) = \text{False}}{\text{Idle}, \, q, \, \sigma \to \text{Idle}, \text{rotate}(q), \, \sigma} \quad \text{ScheduleGuardWait}$$

■ **Figure 7** SOS scheduler semantics for guarded commands

This simplicity, however, is deceptive, because the number of rules in the SOS-style semantics depends on the number of statement kinds *and* the scheduling policy. For example, if one would add the guarded commands of Section 6 to `CoopWhile`, in LAGC it suffices to simply add rule (9) and check that the fairness proof still goes through, while in SOS *three* new rules (Figure 7) are required. Put differently, whenever a scheduling-sensitive statement is added, then in LAGC it suffices to look at the scheduling rules, whereas in SOS one needs to look at each rule for each such statement.

It is possible to write an SOS semantics that emulates the LAGC style if local evaluation and composition rules are carefully separated. In this case, proofs and rules are guided by syntax instead of events, the configurations appearing in rules need to be decomposed in a more complex way, involving meta-notational effort. While it would be possible to represent our approach in SOS-style, the LAGC semantics drastically simplifies the formal arguments. Given the complexity inherent to fairness proofs (see App. E), this is of essence.

## 9 Related Work

**Fairness** Many existing works that deal with fair scheduling reason on the reachable actions of a system [21]. Here we are interested in the specification of fair schedulers at the level of programming languages, which requires to reason more generically, independently of the implemented system. This section describes related work that either has a strong focus on generic fairness proofs, or on target programming languages related to cooperative scheduling.

Daum et al. [9] prove strong fairness of a microkernel scheduler with priorities in Isabelle/HOL. Progress is expressed at a low level (pid of the evolving process), not at a high-level (small-step) semantics such as LAGC. Strong fairness is ensured except when a high-priority process can starve lower priority ones. Apt et al. [3, Chapter 12], following [4, 5], define a universal fair scheduler for pairs $(E, i)$ of enabled ($E$) and selected ($i$) actions and prove it to be fair. Unlike in our approach, the scheduler is connected to a concrete language by way of program transformation that enforces fairness, not via the semantics. Our method allows to study fairness at a semantic level and removes the need for program transformation. Similarly as in our framework, where scheduling decisions are localized in the composition rules of the semantics, one line of work [11, 18] provides abstract local characterizations of fairness via predicate transformers, but does not connect to concrete schedulers. Cooperative scheduling is not considered in any of the papers cited in this paragraph.

In contrast, Muller et al. [24] design and prove responsiveness for a probabilistic scheduler of cooperative threads with priorities. The work is based on the cooperative suspension of threads but they interrupt threads that do not suspend; this prevents assumptions on the finiteness of local computation but compromises the guarantees of cooperative scheduling. Compared to us, beyond the distinction between probabilistic fairness and deterministic scheduling. the main difference is that we are able to relate scheduling to program semantics instead of threads.

A notion of quiescent state was defined for I/O automata [26, 27], where it denotes states at which the automaton only expects inputs. One can define a corresponding fairness notion where I/O automata either reach a quiescent state or loop forever. Our notion of quiescent state shares similarities: in both cases, in a quiescent state a non-deterministic scheduling or communication decision is to be taken, but our notion of fairness is stronger, because it guarantees that each task that can run will run eventually.

**Semantics**    The present article is based on two different ways to express programming language semantics. Small step SOS is the best established and most straightforward approach to semantics for concurrent systems. In contrast, the LAGC semantics is still under development, but it offers an interesting trade-off between compositionality and reasoning on traces. Many other semantic approaches exist, we review some of them in the light of the goals of the present paper.

Evaluation contexts [15] are used to make operational semantics easier to read and more modular. This works by allowing the rules to focus on the reduced terms and by separating rules that express how to reduce inside a context from rules that have an operational effect. We do not use this mechanism here, because the CoopWhile language is too small to benefit from the modularity provided by reduction contexts, even though it would be perfectly possible. On the other hand, we do not think that reduction contexts are useful in the LAGC semantics, because it realizes modularity in a different way.

Modular semantics [23] is an update on SOS aiming at modularity. Mosses observes that contextual elements like store, stack or other elements usually appear in a *runtime configuration* at the source of the target of the reduction when writing a SOS rules. This aspect is highly non-modular. The idea behind modular SOS is to put those elements inside the labels of the reduction relation. The configurations then consist only of programs to be evaluated and rules do not have to be extended when the runtime configuration is extended. It is, however, necessary to restrict how labels can be composed to form a valid trace. Mosses proposes a categorical foundation to formalize this aspect. Dealing with complex concurrent semantics, where several parts of the programs are reduced at the same time or in an interleaved manner, seems to be substantially more complex in modular SOS that in LAGC which has been designed for this very purpose. Specifically, LAGC features symbolic variables to deal with non-evaluated expressions which helps the modular design of the semantics. In fact, abstract states with symbolic variables could be used to make "modular SOS" more modular as well.

Interaction trees [29] are a modular and compositional semantics based on a denotational approach. The semantics produces a tree where nodes are events to represent the interactions of a given program (for example, memory reads and writes). Interaction trees were extended to deal with non-deterministic events and concurrency [8] with an application to the semantics of cooperative scheduling. Abadi and Plotkin [1] define a denotational semantics based on partial traces and use the memory state explicitly to define the composition of traces. However, the denotational perspective sets these kinds of semantics apart from the small step, trace-based approach in LAGC. Even such more modular denotational semantics as these fail to surpass the degree of LAGC's modularity when it comes to cooperative scheduling. This is mostly due to the difficulty to assign semantics to a program with cooperative threads while not all threads are known. In addition, none of the mentioned approaches tackle the problem of fair scheduling.

## 10 Conclusion and Future Work

Our paper makes two *novel*, major contributions to the state of art in the analysis of scheduling of concurrent programs:

1. *A framework to formalize schedulers and fairness based on the programming language semantics.*

   Our LAGC semantics separates local evaluation of sequential computations and their composition into global traces. This permits to define schedulers as a refinement of trace composition rules, leaving the local statement semantics completely untouched. This modularity and the semantic approach, makes the framework adaptable to other programming languages and concurrency paradigms. The resulting scheduling rules are designed in a constructive manner, such that a scheduler implementation can be immediately derived from them. The model is precise enough to allow for formal proofs of fairness. Together, this gives the prospect of a general method for the design of formally proven fair schedulers.

2. *A novel fairness notion adapted to cooperative scheduling.*

   We gave an in-depth, technical study of fairness in the context of cooperative scheduling and the novel concept of *quiescent fairness*, a fairness criterion adapted to that setting. This study is instantiated both with an SOS semantics and with the more recent LAGC semantics. We suggest it as the most suitable fairness criterion for programs with cooperative scheduling. This articles also provides the first formal fairness proof in the context of cooperative scheduling.

Obvious future work is to transfer our approach to other concurrent settings and to implement it for cooperative scheduling languages [6]. In particular, this article should be the formal basis for the future implementation of a scheduler for the ABS language. One could also investigate other trace (hyper-)properties beyond fairness.

**Acknowledgements**   We thank the reviewers for their careful reading of our submission and the valuable feedback. The main part of the presented work was done during a research stay of Hähnle at LIP in Lyon. The support from ENS Lyon in form a visiting professorship and from TU Darmstadt in form of a granted sabbatical is gratefully acknowledged.

## A   Queues

We use queues to express concrete scheduling policies. We define queues with the following standard constructors, functions, and axioms.

**Definition 14** (Queue). *We define $q ::= $ Queue of $T$ as the type of* queues *over an element type $T$. Queue operations are:*

create() — *creates an empty queue*
empty($q$) — *is true iff $q$ is the empty queue*
push($q$, $k$) — *returns a queue $q'$ that is $q$ extended with $k$ as its final element*
pop($q$) — *returns a pair $(q', k)$, where $k$ is the first element in $q$ and $q'$ the remaining queue after $k$ is removed; the operation is undefined iff $q$ is empty*

*The operations are defined by the following (standard) axioms:*

empty(create()),   pop(create()) *is undefined*

$\forall q, k \, \neg$empty(push($q, k$))

$$\forall q, k, q', k' \, \text{pop}(\text{push}(q, k)) = \begin{cases} (\text{create}(), k) & \text{if empty}(q) \\ (\text{push}(q', k), k') & \text{otherwise, where } (q', k') = \text{pop}(q) \end{cases}$$

Assume that the element type of a queue $q$ are continuations, i.e. elements of the form K($s$). Then the operator M returns the *multiset* of non-empty continuations inside a given queue:

$$\text{M}(q) = \begin{cases} \{\} & \text{if empty}(q) \\ \{k\} \uplus \text{M}(q') & \text{if } (q', k) = \text{pop}(q) \wedge k \neq \text{K}(⦰) \\ \text{M}(q') & \text{otherwise, where } (q', k) = \text{pop}(q) \end{cases} \tag{25}$$

**Lemma 1** (Queue properties). *The main properties we use for queues are the following.*

1. *If a non-empty continuation can be popped then it is in* M($q$):
   pop($q$) $= \big((q', \text{K}(s)) \, \wedge \, s \neq ⦰\big) \implies \text{K}(s) \in \text{M}(q)$.

2. *After pushing an element, it is in* M($q$): M(push($q$, K($s$))) $= \text{M}(q) \uplus \{\text{K}(s)\}$.

3. *If pop returns an empty continuation, it does not change* M($q$):
   pop($q$) $= (q', \text{K}(⦰)) \implies \text{M}(q) = \text{M}(q')$.

## B  Scheduler for Procedure Calls Is Deterministic

We state and prove determinacy of the scheduler presented in Section 5.2. It assumes determinacy and completeness of expression evaluation (which in a reasonable setting can be ensured by typing).

*Proof Sketch of Prop. 1.* We proceed by induction on the number of rule applications to obtain $sh$. The induction hypothesis is the claim, strengthened by the following statement: there is at most one unresolved procedure call event in $sh$, i.e. rule (7) is either not applicable or for at most one procedure execution up to renaming.

We need to derive from the determinacy and correctness of expression evaluation that the local evaluation rules for statements are always applicable. For the conditional this follows from the complementary path conditions. For the other rules it is trivial.

<div align="right">□</div>

From Prop. 1 and the fact that all trace composition rules preserve well-formedness of $sh$ we obtain that for a finitely terminating program, exhaustive rule application ends in a state of the form "$sh, q$" with empty($q$).

Determinacy follows from the semantics of the sequential While language which ensures that when $\mathrm{val}_\sigma(s)$ returns several results not equal modulo renaming of variables, then their path conditions are always disjoint (see if statement). If the sequential language were not deterministic the scheduled language would not be either, even though the scheduler itself does not introduce non-determinism.

## C  Proof of Weak Fairness for Spawn – Thm. 1

We first define scheduling distance, then state a lemma relating to this distance and our scheduler.

**Definition 15** (Scheduling Distance). *We define the* position *of a continuation* $\mathrm{K}(s)$ *within* $q$:

$$\mathrm{pos}(q, \mathrm{K}(s)) = \begin{cases} 1 & \textit{if } \mathrm{pop}(q) = (q', \mathrm{K}(s)) \\ 1 + \mathrm{pos}(q', \mathrm{K}(s)) & \textit{if } \mathrm{pop}(q) = (q', \mathrm{K}(s')) \wedge s \neq s' \\ \textit{Undefined} & \textit{if } \mathrm{K}(s) \notin \mathrm{M}(q) \end{cases}$$

*Furthermore, we define the number of unmatched call events in a trace $sh$ as:*

$$\mathrm{unmatched}(sh) = \sum_{m,v} (\#_{sh}(callEv(m, v)) - \#_{sh}(callREv(m, v))) \ .$$

*For any configuration $sh$ with queue $q$, continuation $\mathrm{K}(s) \in \mathrm{M}(q)$, we define the* distance *from scheduling of $\mathrm{K}(s)$ in $q$ as the pair $d(sh, \mathrm{K}(s), q) = (\mathrm{pos}(q, \mathrm{K}(s)), \mathrm{unmatched}(sh))$. We equip this pair with the lexicographic order. If $\mathrm{K}(s) \notin \mathrm{M}(q)$ the distance is undefined.*

We have $d(sh, \mathrm{K}(s), q) \geq (1, 0)$, whenever $\mathrm{K}(s) \in \mathrm{M}(q)$ and, whenever $d(sh, \mathrm{K}(s), q) = (1, 0)$, then the next applicable scheduling rule is rule (6), where $pop(q) = (\mathrm{K}(s), q')$ for some $q'$. The following lemma is proven by simple case analysis:

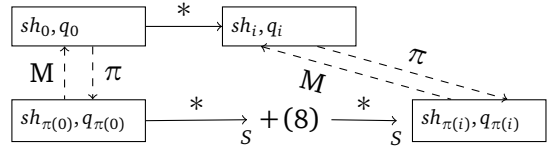

■ **Figure 8**  Relation between non-scheduled and scheduled executions

**Lemma 2** (Scheduling Distance). *Consider a scheduled execution* $sh_0, q_0 \rightarrow_S sh_1, q_1 \rightarrow_S \cdots \rightarrow_S sh_n, q_n \rightarrow_S \cdots$. *For all $s$ and $l \geq 0$ exactly one of the three following cases applies:*

1. $d(sh_l, \mathrm{K}(s), q_l) > (1,0)$ *and* $d(sh_l, \mathrm{K}(s), q_l) > d(sh_{l+1}, \mathrm{K}(s), q_{l+1})$;
2. $d(sh_l, \mathrm{K}(s), q_l) = (1,0)$ *and rule* (6) *is applicable with* $\mathrm{pop}(q_l) = (q_{l+1}, \mathrm{K}(s))$;
3. $s \notin \mathrm{M}(q_l)$.

*Proof of Thm. 1.* Consider a scheduled execution $\langle \sigma_\epsilon \rangle, q_{ini} \rightarrow_S^* sh, q \rightarrow_S^* \cdots$. First we show existence of a derivation of the form $\langle \sigma_\epsilon \rangle, \{\mathrm{K}(s_{main})\} \rightarrow^* sh, M(q) \rightarrow^* \cdots$, then we show that it is weakly fair.

*Existence.* The proof proceeds by induction on the number $n$ of steps in $\rightarrow_S^*$. Assume we obtained $sh, q$ in $n$ many $\rightarrow_S^*$ steps and extend it as $sh, q \rightarrow_S^* sh', q'$. The induction hypothesis gives us the existence of a $\rightarrow^*$-trace until $sh, M(q)$. We perform a case distinction on the kind of rule that was applied to obtain $sh', q'$.

**Rule** (6): In this case $s \neq \emptyset$, hence $\mathrm{K}(s) \in M(q)$ by Lemma 1, so rule (4) is applicable to $M(q)$ and yields $sh', M(q) \uplus \mathrm{K}(s')$. Using the conclusion of rule (6) and Lemma 1 we obtain $sh, M(q) \rightarrow sh', M(q')$.

**Rule** (7): The rule's premises are identical to those of rule (5), so the latter is applicable to $M(q)$ and yields $sh', M(q) \uplus \{\mathrm{K}(s[x \leftarrow y])\}$. Using the conclusion of rule (7) and Lemma 1 we obtain $sh, q \rightarrow sh', M(q')$.

**Rule** (8): The rule gives directly $sh = sh'$ and we obtain $M(q) = M(q')$ from its leftmost premise by Lemma 1. Hence, $sh, M(q)$ is already the desired $\rightarrow$-configuration. Note that empty continuations are automatically simplified in the $\rightarrow$ semantics.

*Mapping.* From the analysis above, it is obvious that one can define an injective mapping $\pi : \mathbb{N} \rightarrow \mathbb{N}$ that gives the $\pi(i)$-th step in $\rightarrow_S$ that corresponds to the $i$-th step in $\rightarrow$ (see illustration in Figure 8):

$$\pi(i) = i + \#(\text{applications of rule (8) to obtain } sh_i, q_i)$$

*Fairness.* Given a rule application sequence $sh_0, q_0 \rightarrow_S sh_1, q_1 \rightarrow_S^* sh_n, q_n \rightarrow_S \cdots$ in the scheduler we show that $sh_{\pi(0)}, \mathrm{M}(q_{\pi(0)}) \rightarrow sh_{\pi(1)}, \mathrm{M}(q_{\pi(1)}) \rightarrow^* sh_{\pi(n)}, \mathrm{M}(q_{\pi(n)}) \rightarrow \cdots$ is weakly fair, i.e.:

$$\forall m, s. \left( \forall n \geq m. \left( \mathrm{K}(s) \in \mathrm{enabled}(sh_{\pi(n)}, \mathrm{M}(q_{\pi(n)})) \right) \Rightarrow \exists n' > m. \left( \mathrm{K}(s) \in \mathrm{M}(q_{\pi(n')}) \right) \right) .$$

Let $m, s$ be arbitrary and assume the left part of the implication holds. We proceed by case analysis on the rule applied to obtain $\mathrm{K}(s) \in \mathrm{enabled}(sh_{\pi(m)}, \mathrm{M}(q_{\pi(m)}))$. Since $\mathrm{enabled}(sh_{\pi(m)}, \mathrm{M}(q_{\pi(m)}))$ is a multiset, it might contain several identical continuations $\mathrm{K}(s)$ to which different rules are applied—it does not matter which one is taken, because our notion of fairness is based on the selected continuation, not on rule selection. However there must be at least one rule applicable, we do a case analysis on that rule. We start with rule (5), because this is the simpler case.

**Rule** (5): in this case, rule (7) is applicable, because it has the same premises as rule (5). We obtain $sh_{\pi(m)}, q_{\pi(m)} \rightarrow_S sh_{\pi(m)+1}, q_{\pi(m)+1}$, where $q_{\pi(m)+1} = \text{push}(q_{\pi(m)}, \text{K}(s))$. By definition of $\pi$ we have $\pi(m+1) = \pi(m) + 1$ and with Lemma 1 we obtain $\text{K}(s) \in \text{M}(q_{\pi(m+1)})$. Hence the fairness condition is achieved for $n' = m + 1$.

**Rule** (4): In this case there is a $s' \in \text{M}(q_{\pi(m)})$ such that $\sigma = last(sh_{\pi(m)})$ and $pc \triangleright \tau \cdot \text{K}(s) \in val_\sigma(s')$. There are two subcases:

(a) Rule (6) is *directly* applicable on the desired statement, i.e. $pop(q_{\pi(m)}) = \text{K}(s')$ . We obtain $sh_{\pi(m)}, q_{\pi(m)} \rightarrow_S sh_{\pi(m)+1}, q_{\pi(m)+1}$, where $q_{\pi(m)+1} = \text{push}(q_{\pi(m)}, \text{K}(s))$, but now $\text{K}(s)$ derives from evaluation of the continuation of $s'$ in $sh_{\pi(m)}$. The reasoning is exactly as above.

(b) The problematic case arises when rule (6) is not directly applicable, either because $\text{pop}(q_{\pi(m)}) \neq \text{K}(s')$ or rule (5) is applicable. In this case we rest our argument on the scheduling distance of $s'$ in $q$ being strictly greater than $(1, 0)$. Successive application of scheduling rules eventually must reach a point, where the distance is $(1, 0)$ and the next applicable scheduling rule is rule (6), where $pop(q) = (\text{K}(s), q')$ for some $q'$. This follows from Lemma 2 applied to the scheduling execution starting at $sh_{\pi_m}, q_{\pi_m}$ and by noticing that $\text{K}(s')$ is persistent in $q_l$ for $l \geq \pi(m)$ except when rule (6) is *directly* applicable. The case of Lemma 2.1 cannot be encountered indefinitely, because the ordering is well-founded, the case of Lemma 2.3 is excluded by definition of weak fairness, so the case of Lemma 2.2 must apply at some $l$, and at this point rule (6) is applicable, with the same result as rule (4) and reasoning as in case (a) above. □

To summarize, the proof of fairness for our scheduler relies on the exhibition of a distance measure that is bounded and decreases systematically. At each point, either the distance reduces or the desired continuation is produced. The same proof principle will be used in the other fairness proofs below.

### D  Proof of Weak Fairness for Guard

We provide a detailed proof of the weak fairness criterion for the scheduler of Guard. As outlined in Section 6, we need to disambiguate syntactically identical statements from differing code contexts, formally:

$$val_\sigma(s_1) = pc \triangleright \tau \cdot \text{K}(s_1') \wedge val_\sigma(s_2) = pc' \triangleright \tau' \cdot \text{K}(s_2') \wedge s_1 \neq s_2 \implies s_1' \neq s_2' \tag{26}$$

We need a similar requirement concerning procedure bodies. The continuation obtained by the valuation of a statement must be different from any procedure body:

$$val_\sigma(s) = pc \triangleright \tau \cdot \text{K}(s') \implies \nexists m. m(x)\{s'\} \in \overline{M} \tag{27}$$

It is sufficient to tag each statement with a line number in the original program to be able to distinguish two statements, so this not a restriction in practice.

We now focus on the adaptation of the proof of Theorem 1 to the extended setting. We show that in the context of *weak fairness* rotations that increase the scheduling distance cannot occur. The central insight is this: if the left part of the implication in the weak fairness Def. 5 is fulfilled and neither rule (7) nor rule (8) are applicable, then also the rotation rule (10) is inapplicable and, in fact, case (a) in the proof of Thm. 1 applies. This is, because selection of rule (6) or (10) is determined by whether the path condition is consistent. But when rule (10) is applied the path condition for the guard is false which contradicts the requirement for weak fairness. Lemma 2 must be adapted, because now rule (10) may be applicable when a statement cannot be scheduled.

**Lemma 3** (Scheduling Distance). *Case 2 of Lemma 2. is restated as follows:*

*2. $d(sh_l, K(s), q_l) = (1, 0)$ and rule (6) or (10) is applicable with $pop(q_l) = (K(s), q_{l+1})$.*

Observe that when rule (10) is applied the scheduling distance of $s$ in $q_l$ increases.

*Proof of fairness for scheduling with guarded commands.* Following the organization of the proof of Thm. 1, we separate existence and fairness.

*Existence.* The application of the rotation rule (10) gives rise to a new case that is completely analogous to that of rule (8). The rest of the existence proof is unchanged.

*Mapping.* The definition of $\pi$ is adapted accordingly to the small change above:

$$\pi(i) = i + \#(\text{applications of rule (8) or (10) to obtain } sh_i, q_i)$$

*Fairness.* Concerning fairness a similar analysis is performed. The cases for rules (5) and (4)(a) are unchanged. Case (4)(b) becomes slightly more complex:

Let $K(s) \in \text{enabled}(sh_{\pi(m)}, M(q_{\pi(m)})$ with $s' \in M(q_{\pi(m)})$ such that $\sigma = last(sh_{\pi(m)})$ and $pc \triangleright \tau \cdot K(s) \in \text{val}_\sigma(s')$. As before we can reason on the scheduling distance for $s'$ in $q_{\pi(m)}$. Due to the change in Lemma 3, it is not always the case that the scheduling distance strictly decreases, but when it increases the premise for weak fairness is not fulfilled.

Indeed, first suppose that for all $l > m$, $K(s) \in \text{enabled}(sh_{\pi(l)}, M(q_{\pi(l)}))$. This implies that rule (4) is applicable with statement $s'$ and consistent path condition, because assumption (26) prevents it from obtaining $K(s)$ via rule (4) from another statement than $s'$, assumption (27) prevents it from obtaining $s$ via rule (5) and a *callREv(m, v)* event for some $m$.

By the same argument as in the proof of Thm. 1, one must obtain after a finite number of steps $d(sh_l, K(s), q_l) = (1, 0)$. At this point, either the path condition is inconsistent and $K(s) \notin \text{enabled}(sh_{\pi(l)}, M(q_{\pi(l)}))$ (because of the analysis of the previous paragraph) which contradicts the weak fairness hypothesis; or statement $s'$ is scheduled and continuation $K(s)$ required for weak fairness is obtained. □

In contrast to Section 5.2, the requirements for strong fairness are not satisfied here. Indeed, Example 8 would progress if strong fairness were achieved, but it loops infinitely with the proposed scheduler.

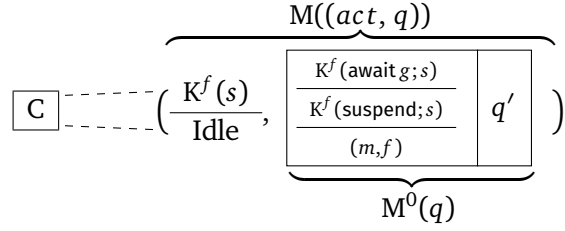

**Figure 9** The relation between continuations in the non-deterministic semantics and the scheduler.

## E   Fairness of Cooperative Scheduling

The more complex structure of configurations in the scheduler rules is reflected in the translation function M from $\rightarrow_S$-traces to $\rightarrow$-traces that occurs in the statement of Thm. 2. The relation is illustrated by the diagram in Figure 9: A scheduler configuration C contains a currently active task $K^f(s)$ or the marker Idle, if no task is currently executed. The task list also contains a queue $q$ of waiting tasks $K^f(\text{await } x; s)$, $K^f(\text{suspend}; s)$, and task execution markers $(m, f)$. We extend the translation function M of equation (25). The auxiliary mapping $M^0$ extracts a set of waiting tasks from the task queue $q$ by ignoring the task execution markers, which have no counterpart in the non-deterministic semantics. Mapping M adds the active task, if there is one.

$$M^0(q) = \begin{cases} \{\} & \text{if empty}(q) \\ \{k\} \uplus M^0(q') & \text{if } (q', k) = \text{pop}(q) \wedge k \neq (m, f) \\ M^0(q') & \text{otherwise, where } (q', k) = \text{pop}(q) \end{cases}$$

$$M((act, q)) = \begin{cases} M^0(q) & act = \text{Idle} \vee act = K^f(\Box) \\ M^0(q) \uplus \{act\} & \text{otherwise} \end{cases}$$

We also define a translation for the traces by simply removing *doneEv* events:

$$M(\varepsilon) = \varepsilon$$
$$M(\tau \frown doneEv) = M(\tau)$$
$$M(\tau \frown t) = M(\tau) \frown t \qquad \text{when } t \neq doneEv$$

The following lemma relates configurations in $\rightarrow$ and $\rightarrow_S$:

**Lemma 4.** *Let* C *occur in a* $\rightarrow_S$*-trace.*

1. *If* $C = (K^f(s), q)$ *for some* $f$, $s \neq \Box$ *and* $q$, *then* $M(C) = \{K^f(s)\} \uplus M^0(q)$.

2. *If* $C = (K^f(\Box), q)$ *for some* $f$ *or* $C = (\text{Idle}, q)$ *and* $\neg\text{empty}(q)$, *then* $M(C) = M^0(q)$.
   *In this case, there is* $q'$, $f'$, *and* $s$ *such that* $\text{pop}(q) = (q', K^{f'}(s))$ *for some suspended statement* $s$ *and* $M^0(q) = \{K^{f'}(s)\} \uplus M^0(q')$.

The scheduling distance for cooperative scheduling needs to take into account a feature not present in Def. 15: the explicit $\rightarrow_S$-step needed for suspending a task that cannot be activated.

**Definition 16** (Scheduling Distance: Cooperative Scheduling). *For any configuration sh, C and continuation* $K^f(s) \in M(C)$ *we define the* distance from scheduling *of* $K^f(s)$ *as the lexicographically ordered triple*

$$d(sh, K^f(s), C) = (\text{pos}_1(C, K^f(s)), \mathit{Idle}(C), \text{unmatched}(sh)) \ , \ \textit{where:}$$

$$\text{pos}_1(C, K^f(s)) = \begin{cases} 1 & \text{if } C = (K^f(s), q) \wedge \neg\text{suspended}(s) \\ \mathit{length}(q) + 2 & \text{if } C = (K^f(s), q) \wedge \text{suspended}(s) \\ 1 + \text{pos}_0(q, K^f(s)) & \text{if } C = (K^f(s'), q) \wedge s \neq s' \\ \text{pos}_0(q, K^f(s)) & \text{if } C = (\mathit{Idle}, q) \end{cases}$$

*Observe that the third line above includes the case when* $s' = \mathbb{0}$.

$$\text{pos}_0(q, K^f(s)) = \begin{cases} 1 & \text{if } \text{pop}(q) = (q', K^f(s)) \\ 1 + \text{pos}_0(q', K^f(s)) & \text{if } \text{pop}(q) = (q', (m, f')) \vee \\ & \quad \big(\text{pop}(q) = (q', K^{f'}(s')) \wedge (s \neq s' \vee f \neq f')\big) \\ \mathit{Undefined} & \text{if } K^f(s) \notin M^0(q) \end{cases}$$

$$\mathit{Idle}((act, q)) = \begin{cases} 1 & \textit{if } act = \mathit{Idle} \\ 0 & \textit{else} \end{cases}$$

$$\text{unmatched}(sh) = \begin{cases} 1 & \textit{if } \text{lastevent}(sh) = \mathit{callEv}(v, m, f) \textit{ for some } v, m, f \\ 0 & \textit{else} \end{cases}$$

Observe that $d(sh, K^f(s), C) \geq (1, 0, 0)$, whenever $K^f(s) \in M(C)$. If $K^f(s) \notin M(C)$ the distance is undefined. Also, if $d(sh, K^f(s), C) = (1, 0, 0)$ then the next applicable scheduling rule is (22). If $d(sh, K^f(s), C) = (1, 1, 0)$ then the next applicable scheduling rule is (20) or (21).

In addition to computing the scheduling distance of a continuation, we also need to compute the scheduling distance of a task activation $(m, f)$. To this end we overload $d$, $\text{pos}_0$, and $\text{pos}_1$ such that we can measure the distance before a task can start in a given configuration. The definition of $\text{pos}_0(q, (m, f))$ is obvious. The definition of $\text{pos}_1$ is:

$$\text{pos}_1(C, (m, f)) = \begin{cases} 1 + \text{pos}_0(q, (m, f)) & \text{if } C = (K^{f'}(s), q) \\ \text{pos}_0(q, (m, f)) & \text{if } C = (\mathit{Idle}, q) \end{cases}$$

**Lemma 5** (Called Procedures). *Consider a* $\to_S$ *execution* $sh_0, C_0 \to_S^* sh, C$. *Assume* $\mathit{callEv}(v, m, f) \in sh$ *and* $\text{lastevent}(sh) \neq \mathit{callEv}(v, m, f)$, *then either* $C = (active, q)$ *and* $(m, f) \in q$ *or* $\mathit{callREv}(v, m, f) \in sh$.

*Proof Sketch.* The proof of this lemma is a straightforward induction on the scheduler rules. Once a procedure is called and the call event has been taken into account, the task activation $(m, f)$ is in the queue until the task is started and the call reaction event is triggered. □

For the purpose of the proof of fairness we introduce an extended reduction relation with a configuration that is augmented with an integer $c \leq 0$, used to count the number of steps between two quiescent states. The absence of infinite local computations ensures that $c$ cannot decrease indefinitely. Now $\rightarrow_S$ has the following signature:

$$sh, C, c \rightarrow_S sh', C', c'$$

In the initial state $c_{init} = 0$. The scheduling rules now update the value of $c$. Rule (18) does not change the value of $c$, i.e. $c' = c$. In rule (22) we have $c' = c - 1$ (we count the number of local steps). In rules (19) and (20) we must reset the counter to $c' = 0$. Finally, for (21), (24), and (23) the value of the counter does not matter, we arbitrarily choose $c' = c$.

The scheduling distance also needs to be extended by one more element in the distance tuple: it is of the form $d(sh, K^f(s), C, c)$ where $c$ is the counter defined above. We overload the definition of the distance so that $d(sh, K^f(s), C, c) = (i, j, k, c)$ whenever $d(sh, K^f(s), C) = (i, j, k)$. The order on the scheduling distance still is the lexical order and the natural integer order on each component.

**Lemma 6** (Bounded Distance). *If a program has no infinite local computation, then for any scheduling computation of this program and any sequence of rule applications in the scheduler $sh_0, C_0, c_0 \rightarrow_S sh_1, C_1, c_1 \rightarrow_S^* sh_n, C_n, c_n \rightarrow_S \cdots$: For any statement $s$ the chain of scheduling distances $d(sh_{\pi(n)}, s, C_{\pi(n)}, c_{\pi(n)})$ does not (strictly) decrease indefinitely. For any task activation $(m, f)$ the chain of scheduling distances $d(sh_{\pi(n)}, (m, f), C_{\pi(n)}, c_{\pi(n)})$ does not (strictly) decrease indefinitely.*

*Proof Sketch.* From the existence proof below we observe there is an $\rightarrow$-execution of the form $M(sh_{\pi(n)}), M(C_{\pi(n)})$ that mimics the scheduling execution. The first three elements of the distance quadruple are positive so they cannot decrease indefinitely.

If there were an infinitely decreasing chain of scheduling distances then there would be $N$, such that from a configuration $sh_N, C_N, c_N$ only $c_N$ decreases indefinitely, i.e. there exist $i, j, k$, such that:

$$\forall m \geq N . d(sh_{\pi(m)}, s, C_{\pi(m)}, c_{\pi(m)}) = (i, j, k, c_{\pi(m)})$$

where $c_{\pi(m)} = c_{\pi(N)} - (m - N)$.

This implies that only rule (22) is applied at each step and that the active continuation is always a non-suspended task. Consequently, for all $m \geq N$, $M(C_{\pi(n)})$ is not a quiescent state which contradicts Definition 3.  $\square$

To proof Thm. 2 we show a more general variant using counters, from which it immediately follows. Let $C^{init} = (K^{f^{init}}(main), \emptyset)$.

**Theorem 3** (Quiescent Fairness with Counter). *Starting from an initial judgment, the scheduler defined by rules (18)–(24) augmented with counters is one possible quiescent fair scheduler of our semantics: If $\langle \sigma_\varepsilon \rangle, C^{init}, c \rightarrow_S sh, C, c' \rightarrow_S \cdots$ then there is a quiescent fair execution $\langle \sigma_\varepsilon \rangle, \{K^{f^{init}}(main)\} \rightarrow^* M(sh), M(C) \rightarrow^* \cdots$ that produces the same trace except for doneEv events.*

*Proof.* As before, the proof consists of two parts: first we show the existence of a derivation of the form $\langle \sigma_\varepsilon \rangle, \{K^{f^{init}}(main)\} \rightarrow^* M(sh), M(C)$, then we show that it is quiescent fair.

*Existence.* The proof proceeds by induction on the number $n$ of steps in $\rightarrow_S^*$. Assume we obtained $sh$, C, $c$ in $n$ many $\rightarrow_S$ steps and extend it as $sh$, C, $c \rightarrow_S sh'$, C', $c'$. The induction hypothesis gives us the existence of a $\rightarrow$-trace until M($sh$), $M$(C). We perform a case distinction on the kind of rule that was applied to obtain $sh'$, C', $c'$.

**Rule** (18): The rule pushes $(m, f)$ on the queue in C, but by definition of $M^0$ this has no effect on M(C') = M(C). Moreover, by definition, M($sh$) = M($sh'$), because they differ only in the *doneEv*. Hence, M($sh$), $M$(C) is already the desired $\rightarrow$-configuration.

**Rule** (19): We can apply rule (17) on the configuration M(C), because the premises are a subset of those of rule (19). This yields M($sh$), ($M$(C) $\uplus$ {K$^f$($s'$)}). Using the conclusion of rule (19) we obtain M($sh$), $M$(C) $\rightarrow$ M($sh'$), M(C') by rule (17) and the fact that $M^0(q) = M^0(q')$.

**Rule** (20): We are in the second case of Lemma 4, all continuations in M(C) are suspended statements (there is no currently active continuation) and one of them is K$^f$($s$) with a consistent path condition, so rule (16) is applicable to M($sh$), M(C). We obtain M($sh$), M(C) $\rightarrow$ M($sh'$), $M$(C') by a direct comparison of rules (16) and (20).

**Rule** (21): Obviously, $M^0(q) = M^0(\text{rotate}(q))$, hence M(C') = M(C), also $sh$ is unchanged. Hence, M($sh$), $M$(C) is already the desired $\rightarrow$-configuration.

**Rule** (22): Here C' = (K$^f$($s'$), $q$). We are in the first case of Lemma 4, M(C) has an active continuation K$^f$($s$), rule (16) is applicable to M($sh$), M(C) and yields M($sh'$), ($M$(C)$\uplus$ {K$^f$($s'$)}). Because rule (22) is applicable and no *doneEv* event can be added to the trace we obtain M($sh$), M(C) $\rightarrow$ M($sh'$), $M$(C') by a direct comparison of rules (16) and (22).

**Rule** (23): The rule gives directly M($sh$) = M($sh'$) and we observe that M(C') = M(C) by definition of $M$ (case where ⦰ is the active continuation), thus M($sh$), $M$(C) is already the desired $\rightarrow$-configuration.

**Rule** (24): By definition of M we have that M(C) = M(C'), also $sh$ is unchanged, thus M($sh$), $M$(C) is already the desired $\rightarrow$-configuration.

*Mapping.* As before, we define an injective mapping $\pi : \mathbb{N} \rightarrow \mathbb{N}$, giving the $\pi(i)$-th step in $\rightarrow_S$ that corresponds to the $i$-th step in $\rightarrow$:

$$\pi(i) = i + \#(\text{applications of rules (18), (21), (24), (23) to obtain } sh_i, C_i, c_i)$$

*Fairness.* Given a sequence of rule applications in the scheduler

$$sh_0, C_0, c_0 \rightarrow_S sh_1, C_1, c_1 \rightarrow_S^* sh_n, C_n, c_n \rightarrow_S \cdots$$

we need to show that the sequence

$$M(sh_{\pi(0)}), M(C_{\pi(0)}) \rightarrow M(sh_{\pi(1)}), M(C_{\pi(1)}) \rightarrow^* M(sh_{\pi(n)}), M(C_{\pi(n)}) \rightarrow \cdots$$

is quiescent fair, i.e.:

$$\forall m, s, f. \Big( \forall n > m. \big( \exists Q. M(C_{\pi(n)}) = Q \Rightarrow K^f(s) \in \text{enabled} \big( M(sh_{\pi(n)}), M(C_{\pi(n)}) \big) \big)$$
$$\Rightarrow \exists n' > m. \big( K^f(s) \in M(C_{\pi(n)}) \big) \Big)$$

Let $m$, $s$, $f$ be arbitrary and assume the left part of the top-most implication holds, i.e. a continuation can be produced. We need to prove that the continuation for statement $s$ is eventually produced, i.e. the scheduler rule producing this statement occurs. We proceed by case analysis on the rule applied to obtain $\mathsf{K}^f(s) \in$ enabled$\big(\mathsf{M}(sh_{\pi(n)}), \mathsf{M}(\mathsf{C}_{\pi(n)})\big)$. Depending on the rule used to produce the statement $s$, either it is the consequence of the evaluation of a statement $s'$ (first case below, rule (16)), or it is the consequence of a task creation rule characterized by $(m, f)$ (second case below, rule (17)). We thus need to prove that either the statement $s$ is produced, or the precondition for quiescent fairness is broken or the scheduling distance associated with the statement $s'$ or the task activation $(m, f)$ decreases. Lemma 6 states that there cannot be an infinite chain of descending scheduling distances, thus $s'$ or $(m, f)$ must eventually be scheduled and the statement $s$ produced.

For each possible rule involved, supposing the precondition for quiescent fairness is not broken and thus the continuation $\mathsf{K}^f(s)$ is reachable, we now merely need to show that one of the two cases applies: (**case 1**) statement $s$ is produced or (**case 2**) the scheduling distance associated with the statement $s'$ or the task activation $(m, f)$ decreases.

**Rule** (16): Recall rule (16) applied to our current configuration $\mathsf{M}(sh_{\pi(n)})$, $\mathsf{M}(\mathsf{C}_{\pi(n)})$:

$$\frac{\mathsf{M}(\mathsf{C}_{\pi(n)}) = Q \uplus \mathsf{K}^f(s') \qquad \sigma = \text{last}(\mathsf{M}(sh_{\pi(n)})) \qquad pc \triangleright \tau \cdot \mathsf{K}(s) \in \text{val}_\sigma^f(s') \qquad \rho \text{ concretizes } \tau \qquad \rho(pc) \text{ consistent} \qquad wf(\mathsf{M}(sh_{\pi(n)}) \ast\!\ast \rho(\tau))}{\mathsf{M}(sh_{\pi(n)}), \mathsf{M}(\mathsf{C}_{\pi(n)}) \rightarrow \mathsf{M}(sh_{\pi(n)}) \ast\!\ast \rho(\tau), Q \uplus \{\mathsf{K}^f(s)\}}$$

In this case a statement is considered for scheduling, but it might be at any position in the queue of the scheduler. By definition of $\mathsf{M}$, the continuation $\mathsf{K}^f(s')$ is in the configuration and the distance $d(sh_{\pi(n)}, \mathsf{K}^f(s'), \mathsf{C}_{\pi(n)}, c_{\pi(n)})$ is well-defined. If lastevent($sh_{\pi(n)}$) is a call event, we first execute rule (18) and consider the configuration $sh_{\pi(n)+1}, \mathsf{C}_{\pi(n)+1}, c_{\pi(n)+1}$ as the starting point of our reasoning instead (rule (18) has no impact on $\rightarrow$, it must be the first executed rule in any case, and after its application the scheduling distance is well-defined). The remaining proof is by case analysis on the reduction $sh_{\pi(n)}, \mathsf{C}_{\pi(n)}, c_{\pi(n)} \rightarrow_S sh_{\pi(n)+1}, \mathsf{C}_{\pi(n)+1}, c_{\pi(n)+1}$.

Rule (19) decreases the scheduling distance (**case 2**):
$d(sh_{\pi(n)}, \mathsf{K}^f(s'), \mathsf{C}_{\pi(n)}, c_{\pi(n)}) = (k, 1, 0, c_{\pi(n)})$, for some $k > 1$ and
$d(sh_{\pi(n)+1}, \mathsf{K}^f(s'), \mathsf{C}_{\pi(n)+1}, c_{\pi(n)+1}) = (k, 0, 0, 0)$.

Rule (20):
$d(sh_{\pi(n)}, \mathsf{K}^f(s'), \mathsf{C}_{\pi(n)}, c_{\pi(n)}) = (k, 1, 0, c_{\pi(n)})$ for some $k$, then either $k = 1$ and the statement $s$ is produced (rule (20) can be applied similarly to rule (16) producing the same statement – **case 1**), or else $k > 1$ and the scheduler may schedule a different task than the one selected by rule (16).

However, we obtain $d(sh_{\pi(n)+1}, \mathsf{K}^f(s'), \mathsf{C}_{\pi(n)+1}, c_{\pi(n)+1}) = (k, 0, 0, 0)$ so the scheduling distance decreases (**case 2**).

Rule (21):
$d(sh_{\pi(n)}, \mathsf{K}^f(s'), \mathsf{C}_{\pi(n)}, c_{\pi(n)}) = (k, 1, 0, c_{\pi(n)})$ for some $k$, it means that $s'$ starts with an await statement followed by statement $s$, and the continuation $\mathsf{K}^f(s)$ must be possible to schedule, i.e. the awaited task has completed. Indeed, we are in a quiescent state, so if $K^f(s)$ were not schedulable, it would not be in enabled$(\ldots)$. Consequentially $k > 1$, and we have $d(sh_{\pi(n)+1}, K^f(s'), C_{\pi(n)+1}, c_{\pi(n)+1}) = (k-1, 1, 0, c_{\pi(n)})$, the scheduling distance of $s'$ decreases (**case 2**).

Rule (22):

$d(sh_{\pi(n)}, K^f(s'), C_{\pi(n)}, c_{\pi(n)}) = (k, 0, 0, c_{\pi(n)})$ for some $k$. Either $k = 1$, in which case we are executing the targeted continuation and $s'$ is not a suspended statement. It is sufficient to compare the premises of rule (22) and rule (16) to conclude (**case 1**). If $k > 1$ then $s'$ is not the statement evaluated by rule (22) and the obtained continuation cannot be $\{K^f(s)\}$. But in this case $d(sh_{\pi(n)+1}, K^f(s'), C_{\pi(n)+1}, c_{\pi(n)+1}) = (k, 0, 0, c_{\pi(n)} - 1)$ and the scheduling distance of $s'$ decreases (**case 2**).

Rule (23):

if $d(sh_{\pi(n)}, K^f(s'), C_{\pi(n)}, c_{\pi(n)}) = (k, 0, 0, c_{\pi(n)})$, then it must be for some $k > 1$ and $d(sh_{\pi(n)+1}, K^f(s'), C_{\pi(n)+1}, c_{\pi(n)+1}) = (k-1, 1, 0, c_{\pi(n)})$, so the scheduling distance of $s'$ decreases (**case 2**).

Rule (24):

Two cases are possible. Either $C_{\pi(n)} = (K^f(s'), q)$ and $s'$ is a suspended statement. In this case $d(sh_{\pi(n)}, K^f(s'), C_{\pi(n)}, c_{\pi(n)}) = (\text{length}(q) + 2, 0, 0, c_{\pi(n)})$ and $d(sh_{\pi(n)+1}, K^f(s'), C_{\pi(n)+1}, c_{\pi(n)+1}) = (\text{length}(q) + 1, 1, 0, c_{\pi(n)})$, so the scheduling distance of $s'$ decreases (**case 2**).

Otherwise, we obtain $d(sh_{\pi(n)}, K^f(s'), C_{\pi(n)}, c_{\pi(n)}) = (k, 0, 0, c_{\pi(n)})$ with $1 < k < \text{length}(q)+2$ and, consequently, $d(sh_{\pi(n)+1}, K^f(s'), C_{\pi(n)+1}, c_{\pi(n)+1}) = (k-1, 1, 0, c_{\pi(n)})$, the scheduling distance decreases (**case 2**).

**Rule** (17): In this case a new task can be spawned. We show that rule (19) can be applied on the correct procedure. First we recall rule (17) applied to our current configuration $M(sh_{\pi(n)})$, $M(C_{\pi(n)})$:

$$\frac{\begin{array}{c} m(x)\{s\} \in \overline{M} \\ \sigma = \text{last}(M(sh_{\pi(n)})) \quad f \in TId \quad pc \triangleright \tau \cdot K(s') \in \text{val}_\sigma^f(m(x)\{s\}) \\ \rho \text{ concretizes } \tau \quad \rho(pc) \text{ consistent} \quad wf(sh_{\pi(n)} ** \rho(\tau)) \end{array}}{M(sh_{\pi(n)}), M(C_{\pi(n)}) \rightarrow M(sh_{\pi(n)}) ** \rho(\tau), M(C_{\pi(n)}) \uplus \{K^f(s')\}}$$

We notice that $\text{val}_\sigma^f(m(x)\{s\})$ starts with a *callREv*$(v, m, f)$ event. Because of well-formedness, we know that the procedure had been called, *callEv*$(v, m, f)$ occurs in $M(sh_{\pi(n)})$, and no *callREv*$(v, m, f)$ event occurs in $M(sh_{\pi(n)})$. If lastevent$(sh_{\pi(n)})$ is a call event, we first execute rule (18) and consider $sh_{\pi(n)+1}, C_{\pi(n)+1}, c_{\pi(n)+1}$ as the starting point of our reasoning instead (rule (18) has no impact on the $\rightarrow$, it must be first executed rule in any case, and after its application the scheduling distance is well-defined).

From Lemma 5 we have that $(m, f)$ is in the queue of suspended tasks and the scheduling distance $d(sh_{\pi(n)}, (m, f), C_{\pi(n)}, c_{\pi(n)})$ is defined and greater or equal to[6] $(1, 1, 0, c_{\pi(n)})$. As before, the remaining proof proceeds by case analysis on the rule application $sh_{\pi(n)}, C_{\pi(n)}, c_{\pi(n)} \rightarrow_S sh_{\pi(n)+1}, C_{\pi(n)+1}, c_{\pi(n)+1}$.

---

[6] Case $(1, 0, 0, c_{\pi(n)})$ is not possible, because $(m, f)$ is not a valid active continuation.

Rule (18): This can only be applied to the creation of a task with a different identifier, and thus we have $d(sh_{\pi(n)}, (m,f), C_{\pi(n)}, c_{\pi(n)}) = (k, i, 1, c_{\pi(n)})$ for some $k \geq 1$ and $i \geq 0$ before its application and $d(sh_{\pi(n)}, (m,f), C_{\pi(n)+1}, c_{\pi(n)+1}) = (k, i, 0, c_{\pi(n)+1})$ with $c_{\pi(n)+1} = c_{\pi(n)}$ after the application of the rule, which decreases the scheduling distance (**case 2**).

Rule (19):
Either $d(sh_{\pi(n)}, (m,f), C_{\pi(n)}, c_{\pi(n)}) = (1, 1, 0, c_{\pi(n)})$ and the right continuation is produced, or the distance is $(k, 1, 0, c_{\pi(n)})$ for some $k > 1$, in which case we have $d(sh_{\pi(n)+1}, (m,f), C_{\pi(n)+1}, c_{\pi(n)+1}) = (k, 0, 0, c_{\pi(n)+1})$ with $c_{\pi(n)+1} = 0$, which decreases the scheduling distance (**case 2**).

Rule (20):
We have $d(sh_{\pi(n)}, (m,f), C_{\pi(n)}, c_{\pi(n)}) = (k, 1, 0, c_{\pi(n)})$ for some $k$, then $d(sh_{\pi(n)+1}, (m,f), Q_{S,\pi(n)+1}, c_{\pi(n)+1}) = (k, 0, 0, c_{\pi(n)+1})$ with $c_{\pi(n)+1} = 0$, the scheduling distance of $(m,f)$ decreases (**case 2**).

Rule (21):
We have $d(sh_{\pi(n)}, (m,f), C_{\pi(n)}, c_{\pi(n)}) = (k, 1, 0, c_{\pi(n)})$ for some $k$, then $d(sh_{\pi(n)+1}, (m,f), C_{\pi(n)+1}, c_{\pi(n)+1}) = (k-1, 1, 0, c_{\pi(n)+1})$ with $c_{\pi(n)+1} = c_{\pi(n)}$, which decreases the scheduling distance of the task activation (**case 2**).

Rule (22):
We have $d(sh_{\pi(n)}, (m,f), C_{\pi(n)}, c_{\pi(n)}) = (k, 0, 0, c_{\pi(n)})$ for some $k$.
There is a currently active task with a non-suspended statement, so procedure $m$ cannot be started, however, the distance decreases: $d(sh_{\pi(n)+1}, (m,f), C_{\pi(n)+1}, c_{\pi(n)+1}) = (k, 0, 0, c_{\pi(n)+1} - 1)$ (**case 2**).

Rule (23):
$d(sh_{\pi(n)}, (m,f), C_{\pi(n)}, c_{\pi(n)}) = (k, 0, 0, c_{\pi(n)})$ for some $k$, and we have $d(sh_{\pi(n)+1}, (m,f), C_{\pi(n)+1}, c_{\pi(n)+1}) = (k-1, 1, 0, c_{\pi(n)+1})$ with $c_{\pi(n)+1} = c_{\pi(n)}$ which decreases the scheduling distance of the task activation (**case 2**).

Rule (24):
There is $k$ such that $d(sh_{\pi(n)}, (m,f), C_{\pi(n)}, c_{\pi(n)}) = (k, 0, 0, c_{\pi(n)})$, then $d(sh_{\pi(n)+1}, (m,f), C_{\pi(n)+1}, c_{\pi(n)+1}) = (k-1, 1, 0, c_{\pi(n)+1})$ with $c_{\pi(n)+1} = c_{\pi(n)}$, the scheduling distance of $(m,f)$ decreases (**case 2**).

$\square$

## About the authors

**Reiner Hähnle** is a Professor of Computer Science at TU Darmstadt, reach him at reiner.haehnle@tu-darmstadt.de. OrcId 0000-0001-8000-7613

**Ludovic Henrio** is a researcher at LIP laboratory in Lyon, reach him at ludovic.henrio@ens-lyon.fr. OrcId 0000-0001-7137-3523